\documentclass{LMCS}

\usepackage{epsf}
\usepackage{graphicx}
\usepackage{url}

\usepackage{amssymb}
\usepackage{graphicx}
\usepackage{enumerate,hyperref}

\newtheorem{THEOREM}{Theorem}[section]
                        {\end{THEOREM}}
\newtheorem{LEMMA}[THEOREM]{Lemma}
                      {\end{LEMMA}}
\newtheorem{COROLLARY}[THEOREM]{Corollary}
\newenvironment{corollary}{\begin{COROLLARY} \hspace{-.85em} {\bf :} }%
                          {\end{COROLLARY}}
\newtheorem{PROPOSITION}[THEOREM]{Proposition}
\newenvironment{proposition}{\begin{PROPOSITION} \hspace{-.85em} {\bf :} }%
                            {\end{PROPOSITION}}
\newtheorem{DEFINITION}[THEOREM]{Definition}
\newenvironment{definition}{\begin{DEFINITION} \hspace{-.85em} {\bf :} \rm}%
                            {\end{DEFINITION}}
\newtheorem{CLAIM}[THEOREM]{Claim}
\newenvironment{claim}{\begin{CLAIM} \hspace{-.85em} {\bf :} \rm}%
                            {\end{CLAIM}}
\newtheorem{EXAMPLE}[THEOREM]{Example}
\newenvironment{example}{\begin{EXAMPLE} \hspace{-.85em} {\bf :} \rm}%
                            {\end{EXAMPLE}}
\newtheorem{REMARK}[THEOREM]{Remark}
                            {\end{REMARK}}

\newcommand{\pro}{\begin{proposition}}
\newcommand{\dfn}{\begin{definition}}
\newcommand{\xam}{\begin{example}}

\newcommand{\ethm}{\end{theorem}}
\newcommand{\elem}{\end{lemma}}
\newcommand{\epro}{\end{proposition}}
\newcommand{\edfn}{\bbox\end{definition}}
\newcommand{\erem}{\bbox\end{remark}}
\newcommand{\exam}{\bbox\end{example}}
\newcommand{\ecor}{\end{corollary}}

\newcommand{\beqn}{\begin{equation}}
\newcommand{\eeqn}{\end{equation}}

\newcommand{\bbox}{\vrule height7pt width4pt depth1pt}
\newcommand{\clm}{\begin{claim}}
\newcommand{\eclm}{\end{claim}}

\newcommand{\boldc}{{\bf c}}

\newcommand{\rimp}{\Rightarrow}

\newcommand{\dimp}{\Leftrightarrow}

\newcommand{\union}{\cup}

\renewcommand{\phi}{\varphi}

\newcommand{\A}{{\cal A}}

\newcommand{\I}{{\cal I}}

\newcommand{\<}{\langle}
\renewcommand{\>}{\rangle}

\renewcommand{\Box}{\mathbin{\vcenter{\hrule
    \hbox{\vrule \kern .6em
          \vbox to .6em{}\vrule}\hrule}}\hspace{.17ex}}

\newcommand{\cf}{cf.~}

\newcommand{\ol}{\setlength{\itemsep}{0pt}\begin{enumerate}}
\newcommand{\eol}{\end{enumerate}\setlength{\itemsep}{-\parsep}}
\newcommand{\ul}{\setlength{\itemsep}{0pt}\begin{itemize}}
\newcommand{\dl}{\setlength{\itemsep}{0pt}\begin{description}}
\newcommand{\edl}{\end{description}\setlength{\itemsep}{-\parsep}}
\newcommand{\eul}{\end{itemize}\setlength{\itemsep}{-\parsep}}

\newcommand\eqdef{=_{\rm def}}
\newcommand{\true}{\mbox{{\it true}}}
\newcommand{\false}{\mbox{{\it false}}}

\newcommand{\commentout}[1]{}

\newcommand{\bi}{\begin{itemize}}
\newcommand{\ei}{\end{itemize}}
\newcommand{\be}{\begin{enumerate}}
\newcommand{\ee}{\end{enumerate}}

\newcommand{\compat}{{\tt \, || \,}}
\newcommand\tbar{\ \equiv\ }

\newenvironment{bogustabbing}{\begin{tabbing}\=\hspace{2em}\=\=\=\kill}%
{\end{tabbing}}

\newenvironment{program*}{%
\begin{bogustabbing}}{\end{bogustabbing}}

\renewcommand{\Box}{\mathbin{\vcenter{\hrule
    \hbox{\vrule \kern .6em
          \vbox to .6em{}\vrule}\hrule}}\hspace{.17ex}}
\renewcommand{\cf}{\bf}
\newcommand{\ff}{\mathit}

\newcommand{\kf}{}

\newcommand{\state}{{\mathit{state}}}
\newcommand{\before}{{\mathit{before}}}
\newcommand{\after}{{\mathit{after}}}
\newcommand{\kind}{{\mathit{kind}}}
\newcommand{\val}{{\mathit{val}}}
\newcommand{\Val}{{\mathit{Val}}}
\newcommand{\AG}{{\mathit{AG}}}

\newcommand{\sval}{{\cf val}}

\newcommand{\rcv}{{\mathit{rcv}}}
\newcommand{\local}{{\mathit{local}}}
\newcommand{\send}{{\mathit{send}}}
\newcommand{\agent}{{\mathit{agent}}}
\newcommand{\first}{{\mathit{first}}}
\newcommand{\pred}{{\mathit{pred}}}
\newcommand{\msg}{{\mathit{msg}}}
\newcommand{\Consistent}{{\mathit{Consistent}}}
\newcommand{\Feasible}{{\mathit{Feasible}}}
\newcommand{\source}{{\mathit{source}}}
\newcommand{\dest}{{\mathit{dest}}}
\newcommand{\initially}{{\mathit{initially}}}
\newcommand{\history}{{\mathit{history}}}
\newcommand{\Dec}{{\mathit{Determinate}}}

\newcommand{\initstate}{\mathit{initstate}}

\newcommand{\esbefore}{{\before}}
\newcommand{\esafter}{{\after}}
\newcommand{\essend}{{\send}}
\newcommand{\esfirst}{{\first}}
\newcommand{\espred}{{\pred}}

\newcommand{\esprec}{{\prec}}
\newcommand{\espreci}{{\prec}_i}
\newcommand{\espreceq}{{\preceq}}

\newcommand{\essucci}{\succ_i}
\newcommand{\essucceqi}{\succeq_i}

\newcommand{\esge}{{\succeq}}
\newcommand{\Act}{\mathit{Act}}

\newcommand{\Pg}{\mathit{Pg}}
\newcommand{\es}{\mathit{es}}

\newcommand{\bs}{\mathbf{s}}
\newcommand{\bsp}{\mathbf{s'}}
\renewcommand{\boldc}{\mathbf{c}}
\newcommand{\Sys}{\mathit{Sys}}
\newcommand{\bSys}{\mathbf{Sys}}
\newcommand{\bCC}{\mathbf{CC}}
\newcommand{\ls}{\mathbf{ls}}

\newcommand{\psat}{\, \hbox{$|$ \hskip -5pt $\approx$ \,}}
\newcommand{\psatI}{\, \hbox{$|$ \hskip -5pt $\approx_I$ \,}}
\newcommand{\psatkbI}{\, \hbox{$|$ \hskip -5pt $\approx_{I}$ \,}}

\newcommand{\psatpr}{\, \hbox{$|$ \hskip -5pt $\approx^{pr}$ \,}}
\newcommand{\psatprnsp}{\hbox{$|$ \hskip -5pt $\approx^{pr}$}}

\newcommand{\quadthree}{\quad \quad \quad}
\newcommand{\quadsix}{\quadthree \quadthree}

\newcommand{\Stable}{{\mathit{Stable}}}
\newcommand{\FairS}{{\mathit{FairSend}}}
\newcommand{\Imp}{{\mathit{Implies}}}
\newcommand{\Rcv}{{\mathit{Rcv}}}

\newenvironment{oldthm}[1]{\par\noindent{\bf Theorem #1:} \em \noindent}{\par}
\newenvironment{oldlem}[1]{\par\noindent{\bf Lemma #1:} \em \noindent}{\par}
\newenvironment{oldcor}[1]{\par\noindent{\bf Corollary #1:} \em \noindent}{\par}
\newenvironment{oldpro}[1]{\par\noindent{\bf Proposition #1:} \em \noindent}{\par}
\newcommand{\othm}[1]{\begin{oldthm}{\ref{#1}}}
\newcommand{\eothm}{\end{oldthm} \medskip}
\newcommand{\olem}[1]{\begin{oldlem}{\ref{#1}}}
\newcommand{\eolem}{\end{oldlem} \medskip}
\newcommand{\ocor}[1]{\begin{oldcor}{\ref{#1}}}
\newcommand{\eocor}{\end{oldcor} \medskip}
\newcommand{\opro}[1]{\begin{oldpro}{\ref{#1}}}
\newcommand{\eopro}{\end{oldpro} \medskip}

\newcommand{\Pgkb}{\Pg^{\mathit{kb}}}
\newcommand{\pgrealizable}{consistent}

\newcommand{\Links}{{\mathit{Links}}}

\newcommand{\snval}{{\cf{val}}}

\newcommand{\idx}{m}

\def\doi{7 (2:14) 2011}
\lmcsheading%
{\doi}
{1--36}
{}
{}
{Feb.~\phantom02, 2009}
{May\phantom.20, 2011}
{}

\begin{document}

\title{Knowledge-Based Synthesis of Distributed Systems Using Event
Structures\rsuper*%
}

\author[M.~Bickford]{Mark Bickford\rsuper a}
\address{Cornell University,Ithaca, NY 14853}
\email{\{markb,rc,halpern,petride\}@cs.cornell.edu}
\thanks{{\lsuper a} Supported in part by AF-AFOSR F49620-02-1-0170.}

\author[R.~Constable]{Robert Constable\rsuper b}
\address{\vskip-6 pt}
\thanks{{\lsuper b} Supported in part by ONR N00014-02-1-0455 and NSF 0208535.}

\author[J.~Y.~Halpern]{Joseph Y.~Halpern\rsuper c}
\address{\vskip-6 pt}
\thanks{{\lsuper{c,d}} Supported in part by NSF
under grants ITR-0325453, CCR-0208535, IIS-0534064,
and  IIS-0812045,
by ONR under
grant N00014-02-1-0455, by the DoD Multidisciplinary University
Research Initiative (MURI) program administered by  ONR under grants
N00014-01-1-0795 and N00014-04-1-0725, and by AFOSR under
grants F49620-02-1-0101 and FA9550-05-1-0055.}

\author[S.~Petride]{Sabina Petride\rsuper d}
\address{\vskip-6 pt}


\titlecomment{{\lsuper*}A preliminary version of this paper appeared in Proceedings of
the 11th International Conference on Logic for Programming,
Artificial Intelligence, and Reasoning LPAR 2004, pp.
449$\mbox{-}$465. }

\begin{abstract}
To produce a program guaranteed to satisfy a given specification one
can  synthesize it from a formal constructive proof that a
computation satisfying that specification exists. This process is
particularly effective if the specifications are written in a
high-level language that makes it easy for designers to specify
their goals. We consider a high-level specification language that
results from adding {\em knowledge}  to a fragment of Nuprl
specifically tailored for specifying distributed protocols, called
{\em event theory}. We then show how high-level {\em knowledge-based
programs} can be synthesized from the knowledge-based specifications
using  a proof development system such as Nuprl. Methods of Halpern
and Zuck \cite{HZ} then apply to convert these knowledge-based
protocols to ordinary protocols. These methods can be expressed as
heuristic transformation tactics in Nuprl.
\end{abstract}

\maketitle

\section{Introduction}\label{sec: intro}
Errors in software are extremely costly and disruptive.
\commentout{NIST (the
National Institute of Standards and Technology) estimates the cost
of software errors to the US economy at \$59.5 billion per year.
}
One approach to minimizing errors is to synthesize programs from
specifications. Synthesis methods have produced highly reliable
moderate-sized programs in cases where the computing task can be
precisely specified. One of the most elegant synthesis methods is
the use of so-called {\em correct-by-construction} program synthesis
(see, e.g., \cite{Ber03,Con71,book,CH88,Coq96,PMW93}.
Here programs are constructed from {\em proofs}
that the specifications are satisfiable. That is, a constructive
proof that a specification is satisfiable gives a program that
satisfies the specification. This method has been successfully used
by several research groups and companies to construct large complex
{\em sequential} programs;
and it has been used to synthesize distributed protocols such as
Paxos \cite{Lam98}, and various authentication protocols (see www.nuprl.org).

The Cornell Nuprl proof development system was among the first tools
used to create correct-by-construction functional and sequential
programs \cite{book}. Nuprl has also been used extensively to
optimize distributed protocols \cite{LKvR+99}, and to formalize them in the
language of I/O Automata \cite{BKVL01}. Recent work by two of the authors
\cite{CB08} has resulted in the definition of a fragment of the
higher-order logic used by Nuprl tailored to specifying distributed
protocols, called {\em event theory}, and the extension of Nuprl
methods to synthesize distributed protocols from specifications
written in event theory \cite{CB08}.
Moreover, the current version of the Nuprl prover is
itself a distributed system
\cite{ABC06}.

However, as has long been recognized \cite{FHMV}, designers
typically think of specifications at a high level, which often
involves knowledge-based statements.   For example, the goal of a
program might be to guarantee that a certain process knows
certain information.  It has been argued that a useful way of
capturing these high-level knowledge-based specifications is by
using high-level {\em knowledge-based programs} \cite{FHMV,FHMV94}.
Knowledge-based programs are an attempt to capture the intuition
that what an agent does depends on what it knows.  For example, a
knowledge-based program may say that process 1 should stop sending a
bit to process 2 once process 1 knows that process 2 knows the
bit.  Such knowledge-based programs and specifications have been
given precise semantics by Fagin et al.~\cite{FHMV,FHMV94}. They
have already met with some degree of success, having been used
 both to
help in the design of new protocols and to clarify the understanding
of existing protocols \cite{DM,HZ,SV02}.

In this paper, we add knowledge operators to event theory raising
its level of abstraction and show by example that knowledge-based
programs can be synthesized from constructive proofs that
specifications in event theory with knowledge operators are
satisfiable. Our example uses the {\em sequence-transmission
problem (STP)}, where a sender must transmit a sequence of bits to a
receiver in such a way that the receiver eventually knows
arbitrarily long prefixes of the sequence. Halpern and Zuck
\cite{HZ} provide
knowledge-based programs for the sequence-transmission problem, prove them
correct, and show that many standard programs for the problem in the
literature can be viewed as implementations of their high-level
knowledge-based
programs.
Here we show that %
one of these
knowledge-based programs can be synthesized from the specifications
of the problem, expressed in event theory augmented by knowledge.
We can then translate the arguments of Halpern and Zuck to Nuprl, to
show that the knowledge-based
program can be transformed to the standard programs in the
literature.
This paper relies heavily on prior work on knowledge-based
programs of Halpern et al.~\cite{FHMV,FHMV94,HZ}; the novelty
lies in offering a proof of concept that knowledge-based specifications
and programs can be formulated in a constructive logic, and that
knowledge-based programs can be synthesized in a semi-automatic system
like Nuprl.

Engelhardt, van der Meyden, and Moses \cite{EMMFoss,EMM01} have
also provided techniques for synthesizing knowledge-based programs
from knowledge-based specifications, by successive refinement.  We
see their work as complementary to ours.   Since our work is based
on Nuprl, we are able to take advantage of the huge library of
tactics provided by Nuprl to be able to generate proofs.  The
expressive power of Nuprl also allows us to formalize all the
high-level concepts of interest (both
epistemic and temporal) easily.  Engelhardt, van der Meyden, and
Moses do not have a theorem-proving engine for their language.
However, they do provide useful refinement rules that can easily be
captured as tactics in Nuprl.

The paper is organized as follows. In the next section we give a
brief overview of the
Nuprl system,
review event theory, discuss the type of programs we use
(distributed message automata), and show how automata can be
synthesized from a specification. In Section~\ref{sec: k} we
review epistemic logic, show how it can be translated into Nuprl,
and show how knowledge-based automata can be captured in Nuprl. The
sequence-transmission problem is analyzed in Section~\ref{sec:stp}.
We conclude with a discussion of related work and
future research in Section~\ref{sec:conc}.

\section{Synthesizing programs from constructive proofs}\label{sec:overview}

\subsection{Nuprl: a brief overview}\label{sec:Nuprl}
Much current work on formal verification using theorem proving,
including Nuprl, is based on type theory (see \cite{ABC06} for a
recent overview). A type can be thought of as a set with structure
that facilitates its use as a data type in computation; this
structure also supports constructive reasoning. The set of types is
closed under the product space and function constructors $\ff{ \times}$
and $\ff{
\rightarrow}$, so that if $\ff{ A}$ and $\ff{ B}$ are types, so are
$\ff{ A \times B}$ and $\ff{ A \rightarrow B}$,  where, intuitively,
$\ff{ A \rightarrow B}$ represents the computable functions from
$\ff{ A}$ into $\ff{ B}$.

{\em Constructive} type theory (also called \emph{computational} type
theory), on which Nuprl is based, was
developed to provide a foundation for computer science and constructive
mathematics. The
key feature of constructive mathematics is that ``there exists'' is
interpreted as ``we can construct (a proof of)''. Reasoning in the
Nuprl type theory is intuitionistic \cite{brouwer23}, in the sense
that proving a certain fact is understood as  constructing evidence
for that fact. For example, a proof of the fact that ``there exists
$\ff x$ of type $\ff A$" builds an object of type $\ff A$, and a
proof of the fact ``for any object $\ff x$ of type $\ff A$ there
exists an object $\ff y$ of type $\ff B$ such that the relation $\ff
R(x,y)$ holds" builds a function $\ff f$ that associates with each
object $\ff a$ of type $\ff A$ an object $\ff b$ of type $\ff B$
such that $\ff R(a,b)$ holds.

\commentout{ Based on the well-known Curry-Howard isomorphism,
formulas can be seen as types. A proof in the intuitionist type
theory  of a formula $\ff \varphi$ is a proof that the type
associated with $\ff \varphi$ is inhabited. For example, a proof of
the formula $\ff {\exists x:A.~ P(x)}$ builds a  pair consisting of
an object $\ff a$ of type $\ff A$ and an object of the type
associated with $\ff{P(a)}$. This logic is called {\em constructive
first-order logic}; constructive higher-order logics can also be
defined, but for the purpose of this paper first-order logic
suffices. The type of all propositions is denoted as $\ff Prop$. }

One consequence of this approach is that
 the principle of excluded middle does not apply: while in classical  logic,
$\ff \varphi \vee \neg \varphi$ holds for all formulas $\ff
\varphi$, in constructive type theory, it holds exactly when we have
evidence for either $\ff \varphi$ or $\ff \neg \varphi$, and we can
tell from this evidence which of $\ff \varphi$ and $\ff \neg
\varphi$ it supports. \commentout{ Clearly, this ought not to be the
case for any formula $\ff \varphi$. Typical examples encode in $\ff
\varphi$ statements as ``the Turing machine M halts", and reduce the
problem of proving $\ff \varphi \vee \neg \varphi$ to the halting
problem, well-known to be undecidable. It is however straightforward
to give a classical interpretation to formulas in the first-order
logic: one only needs to state the principle of excluded middle as
an axiom, and interpret the logical connectives in the standard way.
For the rest of this paper, we call formulas for which the principle
of excluded middle holds $\ff \varphi \vee (\neg \varphi)$ is
constructively true,
as {\em decidable}; we write this as $\ff Decidable(\varphi)$.%
\footnote{This should not be confounded with the standard notion of
decidability: a problem is decidable if there exists an algorithm
for solving it. It will be clear from context when we refer to a
problem as being decidable, and when we refer to a formula as
decidable.} } A predicate $\Dec$ is definable in Nuprl such that
$\Dec(\varphi)$ is true iff the principle of excluded
middle holds for formula $\varphi$.
(From here on in, when we say that a formula is true, we mean that it is
constructively true, that is, provable in Nuprl.)

In this paper, we focus on synthesizing programs from
specifications. Thus we must formalize these notions  in Nuprl. As a
first step, we define a type $\ff{Pgm}$ in Nuprl and take programs
to be objects of type $\ff{Pgm}$.  Once we have defined
$\ff{Pgm}$, we can define other types of interest.
These definitions rely on a formalization of the notion of
executions {\em consistent\/} with a program, that is, executions that
could have been generated by running the program. As will be clear
in the next sections, we can formally define in Nuprl a notion of
consistency for the programs and executions considered in this
paper.
\dfn\label{def:main} A {\em program semantics} is a function $\ff{
S}$ of type $\ff{ Pgm \rightarrow Sem}$ assigning to each  {\em
program} $\ff{ Pg }$ of type $\ff{ Pgm}$ a {\em meaning} of  type
$\ff{Sem}=2^{\ff{Sem}'}$, where $\ff{Sem}'$ is the type of {\em
executions} consistent with the program $\ff{Pgm}$ under the
semantics $\ff{S}$.
A {\em specification} is a predicate $\ff{ X}$ on $\ff{Sem}'$. A
program $\ff Pg$ {\em satisfies} the specification $\ff X$
if $\ff{X(e)}$ holds for all $e$ in $\ff{S(Pg)}$.
A specification $\ff{X}$ is {\em satisfiable} if
there exists a program that satisfies $\ff X$. \edfn

As Definition~\ref{def:main} suggests, all objects in Nuprl are
typed. To simplify our discussion, we typically suppress the type
declarations. Definition~\ref{def:main} shows that the
satisfiability of a specification is definable in Nuprl. The key
point for the purposes of this paper is that from a constructive
proof that $\ff X$ is satisfiable, we can extract a program that
satisfies $\ff{ X}$.

Theoremhood in constructive type theory is highly undecidable, so we
cannot hope to
construct a proof completely automatically.  However, experience has
shown that, by having a large library of lemmas and proof tactics,
it is possible to ``almost'' automate quite a few proofs, so that
with a few hints from the programmer, correctness can be proved. For
this general constructive framework to be useful in practice, the
parameters $\ff{ Pgm}$, $\ff{Sem}'$, and $\ff{ S}$ must be chosen so
that (a) programs are concrete enough to be compiled, (b)
specifications are naturally expressed as predicates over $\ff{
Sem}'$, and (c) there is a small set of {\em rules} for producing
proofs of satisfiability.

To use this general framework for synthesis of {\em distributed},
{\em asynchronous} algorithms, we choose the programs in $\ff{ Pgm}$
to be {\em distributed message automata}. Message automata are
closely related to {\em IO-Automata} \cite{LT} and are
similar to {\em UNITY} programs \cite{CM88} (but with
message-passing rather than shared-variable communication).
 We
describe distributed message automata in Section \ref{sec:dma}. As
we shall see, they satisfy criterion (a).

The semantics of a program is the {\em system}, or set of {\em
runs}, consistent with it. Typical specifications in the literature
are predicates on runs. We can view a specification as a predicate
on systems by saying that a system satisfies a specification exactly
if all the runs in the system satisfy it. To meet criterion (b),
we formalize runs as structures that we call {\em event structures},
much in the spirit of Lamport's \cite{Lamclocks} model of events
in distributed systems. Event structures are explained in more
detail in the next section. We have shown \cite{CB08} that, although
satisfiability is undecidable, there is indeed a small set of rules
from which we can prove satisfiability in many cases of interest;
these rules are discussed in Section~\ref{sec:dma}.

\subsection{Event structures}\label{sec:eventsystems}
Consider a set $\ff{AG}$ of processes or {\em agents}; associated
with each agent $\ff i$ in $\ff{AG}$ is a set $\ff X_i$ of {\em
local variables}. Agent $i$'s local state at a point in time is
defined as the values of its local variables at that time. We assume
that the sets of local variables of different agents are disjoint.
Information is communicated by message passing. The set of links is
$\Links$. Sending a message on some link $\ff{ l} \in \Links$ is
understood as enqueuing the message on $\ff{ l}$, while receiving a
message corresponds to dequeuing the message. Communication is
point-to-point: for each link $\ff{l}$ there is a unique agent $\ff{
\source(l)}$ that can send messages on $\ff{ l}$, and a unique agent
$\ff{ \dest(l)}$ that can receive message on $\ff{ l}$. For each
agent $\ff i$ and link $\ff l$ with $\ff \source(l)=i$, we assume
that $\ff \msg(l)$ is a local variable in $\ff X_i$.
Intuitively, sending a message $m$ will be identified with setting the
variable $\msg(l)$ to $m \ne \bot$.

We assume that communication is asynchronous, so there is no global
notion of time. Following Lamport \cite{Lamclocks}, changes to
the local state of an agent are modeled as {\em events}.
Intuitively, when an event ``happens'', an agent either sends a
message, receives a message or chooses some values (perhaps
nondeterministically). As a result of receiving the message or the
(nondeterministic) choice, some of the agent's local variables are
changed.

Lamport's theory of events is the starting point of our formalism.
To help in  writing concrete and detailed specifications, we add
more structure to events. Formally, an event is a tuple with three
components. The first component of an event $\ff e$ is an agent
${\ff i\in AG}$, intuitively the agent whose local state changes
during event $e$. We denote ${\ff i}$  as $\ff \agent(e)$. The
second component of $\ff e$ is its {\em kind}, which is either a
link $\ff l$ with $\ff \dest(l)=i$ or a local action $\ff a$, an
element of some given set $\ff{Act}$ of local actions. The only
actions in $\ff {Act}$ are those that set local variables to certain
values. We  denote this component   as $\ff \kind(e)$. We often
write $\ff \kind(e)=\rcv(l)$ rather than $\ff \kind(e)=l$ to
emphasize the fact that $\ff e$ is a receive event; similarly we
write $\ff \kind(e)=\local(a)$ rather than $\ff \kind(e)=a$ to
emphasize the fact that $\ff a$ is a local action.
The last component of $\ff e$ is its {\em value} $\ff v$, a tuple of
elements in some domain $\Val$; we denote this component as  $\ff
\val(e)$. If $\ff e$ is a receive event, then $\ff \val(e)$ is the
message received when $\ff e$ occurs; if $\ff e$ is a local event
$a$, then $\ff \val(e)$ represents the tuple of values  to which the
variables are set by $a$. (For more details on the reasons that led
to this formalism, see \cite{EvSys05}.)

Rather than having a special kind  to model send events, we
 model the sending of a message on link
$\ff l$ by changing the value of the local variable $\ff \msg(l)$ that
describes the message sent on $l$. A special value $\ff \bot$
indicates that no message is sent when the event occurs; changing
$\ff msg(l)$ to a value other than $\ff \bot$ indicates that a
message is sent on $\ff l$. This way of modeling send events has
proved to be convenient. One advantage is that we can model
multicast: the event $\ff e$ of $\ff i$ broadcasting a message $\ff
m$ to a group of agents just involves a local action that sets $\ff
\msg(l)$ to $\ff m$ for each link $\ff l$ from $\ff i$ to one of the
agents in the group. Similarly, there may be an action in which
agent $\ff i$ sends a message to some agents and simultaneously
updates other local variables.

Following Lamport \cite{Lamclocks}, we model an execution of  a
distributed program as a sequence of events satisfying a number of
natural properties. We call such a sequence an event structure.%
\footnote{We use the term {\em sequence} as a simplification. As
explained in the remainder
of the section,
just as for Lamport,
executions are technically partial orders on events
respecting local orders
and causality.}
We take an event structure $\ff es$ to be a tuple consisting of a set
$\ff E$ of events and a number of additional elements that we now describe.
These elements include the functions $\ff dest$, $\ff
source$, and $\ff msg$ referred to above, but there are others. For
example, Lamport assumes that every receive event $\ff{e}$ has a
corresponding (and unique) event where the message received at
$\ff{e}$ was sent. To capture this in our setting, we assume that
the description of the event structure $\ff es$ includes a function
$\ff \send$ whose domain is the receive events in $\ff es$ and whose
range is the set of events in $\ff es$; we require that $\ff
\agent(\send(e))=\source(l)$
if $\kind(e)=\rcv(l)$. Note that, since we allow multicasts,
different receive events may have the same corresponding send event.

For each $\ff{ i \in AG}$, we assume that the set of events $\ff{
e}$ in $\ff es$ associated with $\ff i$ is totally ordered. This
means that, for each event $e$, we can identify the  sequence of
events ({\em history}) associated with agent $i$ that preceded $\ff
e$. To formalize this,  we assume that, for each agent $\ff i\in
AG$, the description of $\ff es$ includes a total order $\esprec_i$
on the events e in $\ff es$ such that $\ff \agent(e)=i$. Define a
predicate $\ff \esfirst$ and function $\espred$ such that
$\esfirst(e)$ holds exactly when $\ff{ e}$ is the first event in the
history associated with $\ff{ \agent(e)}$ in $\ff es$; if
$\esfirst(e)$ does not hold, then $\espred(e)$ is the unique
predecessor of $e$ in $\ff es$. Following Lamport
\cite{Lamclocks}, we take  $\ff{ \esprec}$ to be the least
transitive relation on events  in $\ff es$ such that $\ff send(e)
\esprec e$ if $e$ is a receive event and $e \esprec e'$ if $e
\espreci e'$. We assume that $\ff \esprec$ is well-founded. We
abbreviate  $\ff (e' \esprec e)\vee (e=e')$ as $\ff e' \espreceq e$,
or $\ff e \esge e'$. Note that $\esprec_i$ is defined only for
events associated with agent $i$: we write $e\esprec_i e'$ only if
$\agent(e)=\agent(e')=i$.

The  local state of an agent defines the values of all the variables
associated with the agent.
While it is possible that an event structure contains no events
associated with a particular agent, for ease of exposition, we
consider only event structures in which each agent has at least one
local
state, and denote the initial
local
state of agent $i$ as $\initstate_i$.
(Note that one way to ensure this is to assume that each local variable
has an initial value; the initial state is the state that assigns each
local variable its initial value.)
In event structures $\es$
where at least one event associated with a given agent $i$ occurs,
$\initstate_i$ represents $i$'s
local
state before the first event associated with $i$ occurs in
$\es$. Formally, the
\emph{local state} of an agent $i$ is a function
that maps $X_i$ and a special symbol $\snval_i$ to values.
(The role of $\snval_i$
will be explained when we
give the semantics of the logic.) If $x \in X_i$, we write $s(x)$ to
denote the value of $x$ in $i$'s local state $s$.
Similarly, $s(\snval_i)$ is the value of $\snval_i$ in $s$.
If $\agent(e) = i$, we take $\ff{\state~\esbefore~e}$ to be the
local
 state of agent $i$ before $e$; similarly,
$\ff{\state~\esafter~e}$ denotes $i$'s
local
state after event $\ff{e}$ occurs. The value
$(\ff{\state~\esafter~e})(x)$ is in general different from
$(\ff{\state~\esbefore~e})(x)$. How it differs depends on the event
$e$, and will be clarified in the semantics. We assume that
$(\state~\esafter~e)(\snval_i) = \val(e)$; that is, the value of
the special symbol $\snval_i$ in a
local
state is just the value of the event that it follows. If $x \in
X_i$, we take $\ff{ x~ \esbefore~ e}$ to be an abbreviation for
$(\ff{\state~\esbefore~e})(x)$; that is, the value of $x$ in the
state before $e$ occurs;  similarly, $\ff{x~\esafter~ e}$  is
an abbreviation for $(\ff{\state~\esafter~e})(x)$.

\begin{example}\label{xam:events} Suppose that $\ff Act$ contains $\ff \send$
and $\ff \send\mathit{+inc(x)}$, where $\ff x \in X_i$, and that
$\ff Val$ contains the natural numbers. Let $\ff n$ and $\ff v$ be
natural numbers. Then
\begin{enumerate}[$\bullet$]
\item the event of agent $\ff i$ receiving message $\ff m$ on link $\ff l$
in the event structure $\ff es$ is modeled by the tuple  $\ff
e=(i,l,m)$, where $\ff \agent(e)=i$, $\ff \kind(e)=\rcv(l)$, and
$\ff \val(e)=m$;
\item the event of agent $\ff i$ sending message $\ff m$ on link $\ff l$
in $\ff es$ is represented by the tuple $\ff e=(i,\send,m)$, where
$\msg(l)~\esafter~e=m$;%
\item the event $\ff e$ of agent $\ff i$ sending $\ff m$ on link $\ff l$
and incrementing its local variable $\ff x$ by $\ff v$ in $\ff es$
is represented by the tuple $e$ such that $\agent(e) = i$, $\ff
\kind(e)=\send\mathit{+inc(x)}$, and $\ff \val(e)=\langle
m,v'\rangle$, where $\ff \msg(l)~\esafter~e=m$ and $\ff
x~\esafter~e=x~\esbefore~e + v = v'$.
\end{enumerate}
\end{example}

\begin{definition}\label{def:ev structure} An event structure is a tuple 
$$
\begin{array}{ll}
\es = 
&  \langle \mathit{AG}, \Links, \source, \dest, \Act, \lbrace X_i
\rbrace_{i \in \mathit{AG}}, \Val,
\lbrace \initstate_i \rbrace_{i \in \mathit{AG}},
E, \agent,  \\
& \send, \first, \lbrace \prec_i \rbrace_{i\in
\mathit{AG}}, \prec \rangle 
\end{array}
$$
 where $\mathit{AG}$ is a set of
agents, $\Links$ is a set of links such that $\source :$ $\Links
\longrightarrow \mathit{AG}$, $\dest :\Links \longrightarrow
\mathit{AG}$, $\Act$ is a set of actions, $X_i$ is a set of
variables for agent $i \in \mathit{AG}$ such that, for all links $l
\in \Links$, $\msg(l)\in X_i$ if $i=\source(l)$, $\Val$ is a set of
values,
$\initstate_i$ is the initial local state of agent $i \in \AG$,
$E$ is a
set of events for agents $\mathit{AG}$, kinds $\mathit{Kind} =
\Links \cup \Act$, and domain $\Val$, functions $\agent$, $\send$
and $\first$ are defined as explained above, $\prec_i$s are local
precedence relations and $\prec$ is a causal order such that the
following axioms, all expressible in Nuprl, are satisfied:
\begin{enumerate}[$\bullet$]
\item if $\ff{ e}$ has kind $\ff{ \rcv(l)}$, then the value of $\ff e$
is the message sent on $\ff l$ during event $\essend(e)$, $\agent(e)
= \dest(l)$, and $\agent(\essend(e)) = \source(l)$:
$$
\begin{array}{ll}
\ff{\forall e\in es.\forall l}.~ (\kind(e)=\rcv(l)) \Rightarrow\\
 ~~\ff{(\val(e)=\msg(l)~\esafter~\essend(e))}
\wedge {\ff  (\agent(e)=\dest(l))} \wedge {\ff
(\agent(\essend(e))=\source(l))}.
\end{array}
\footnote{For simplicity, in the remainder of this paper, we abuse
notation and write $e\in \es$ to indicate that $e$ is an event that occurs
in $\es$.}
$$
\item for each agent $\ff i$, events associated with $\ff i$ are totally ordered:
$$\forall e \in es. \forall e' \in es. (\agent(e) = \agent(e') = i \rimp
e\espreci e' \lor  e'\espreci e \lor e=e').$$
\item $\ff e$ is the first event associated with
     agent $\ff i$ if and only if there is no event associated with $\ff i$ that precedes $\ff e$:
     $$\ff{\forall e\in es~\forall i. ~(\agent(e)=i)\Rightarrow (
      \esfirst(e) \Leftrightarrow \forall e'\in es. ~\neg (e'
      \espreci e))}.$$
\item the initial
local
state of agent $i$ is the
state before the first event
associated
with $i$, if any:
$$
\begin{array}{l}
\ff{\forall i.~(\forall e\in \es.~ (\agent(e)=i \Rightarrow
     ( \first(e) \Leftrightarrow (\state~\before~e =
     \initstate_i))))}.
\end{array}
$$
\item the predecessor of an event $\ff e$
immediately precedes $\ff e$ in the causal order:
$$
\begin{array}{l}
\ff{\forall e\in es.~\forall i.~((\agent(e)=i) \land \neg
\esfirst(e)) \Rightarrow }\\
\quad\ff{((\espred(e)\espreci e)\wedge (\forall e'\in es. ~\neg
(\espred(e)\espreci e' \espreci e)))}.
\end{array}
$$
\item the local variables of agent $\ff{ \agent(e)}$ do not
change value between the predecessor of $\ff e$ and $\ff{ e}$:
$$
\begin{array}{l}
\ff{ \forall e\in es. ~\forall i.~(
\agent(e)=i  \land \neg \esfirst(e)) \Rightarrow }\\
\quadsix \ff{\forall x\in X_i.~(x ~\esafter~ \espred(e) = x
~\esbefore ~e)}.
\end{array}
$$
\item the causal order $\ff \esprec$ is well-founded:
$$
\ff{\forall P. ~(\forall e.~(\forall e'\esprec e.~P(e'))\rimp
P(e))\rimp (\forall e.~P(e))},$$ where $\ff P$ is an arbitrary
predicate on events.
(It is easy to see that this axiom is sound if $\esprec$ is
well-founded.  On the other hand, if $\esprec$ is not well-founded,
then let $P$ be a predicate that is false exactly of the events $e$
such that there  is an infinite descending sequence starting
with $e$. In this case, the antecedent of the axiom holds, and the
conclusion does not.)
\end{enumerate}
\end{definition}

\noindent In our proofs, we will need to argue that two events $e$ and $e'$
are either causally related or they are not. It can be shown
\cite{CB08} that this can be proved
in constructive logic iff the predicate $\ff \esfirst$ satisfies the
principle of excluded middle. We enforce this by adding the
following axiom to the characterization of event structures:
$$\ff{\forall e\in es.~\Dec(\esfirst(e))}.$$
The set of event structures is definable in Nuprl (see \cite{CB08}).
We use event structures to model executions of distributed systems.
We show how this can be done in the next section.

\subsection{Distributed message automata}\label{sec:dma}
As we said, the programs we consider are \emph{message automata}.
Roughly speaking, we can think of message automata as
nondeterministic state machines, though certain differences exist.
Each basic message automaton is associated with an agent $\ff{ i}$;
a message automaton associated with $i$ essentially says that, if
certain preconditions hold, $i$ can take certain local actions. (We
view $\ff{ receive}$ actions as being out of the control of the
agent, so the only actions governed by message automata are local
actions.) At each point in time, $\ff{ i}$ nondeterministically
decides which actions to perform, among those  whose precondition is
satisfied. We next describe the syntax and semantics of message
automata.

\subsubsection{Syntax}\label{sec:syntax}
We consider a first-order language for tests in automata.
Fix a set $\AG$ of agents, a set
$X_i$ of local variables for each agent $i$ in
$\AG$, and a set $X^*$ of variables that includes $\union_{i \in \AG}
X_i$ (but may have other variables as well).  The language also includes
special constant symbols $\sval_i$, one for each agent $i\in {\ff
AG}$,
predicate symbols in some finite set ${\it P}$, and
function symbols in some finite set ${\it F}$.
Loosely speaking, $\sval_i$ is used to denote the value
of an event associated with agent $i$;
constant symbols other than
$\sval_1, \ldots, \sval_n$ are just $0$-ary function symbols in $\it F$.
We allow quantification only over variables other than local
variables; that is, over variables $x\notin \cup_{i\in \AG} X_i$.
Allowing non-local variables is not an artificial generalization;
just by looking at a few classic distributed problems, we can see that
non-local variables are ubiquitous.
For example, in a problem where each agent has an input
variable and the goal is for agents
to compute an aggregate of the local inputs, the aggregate is a
non-local variable.

Message automata are built using a small set of {\em basic
programs}, which may involve formulas in the language above.
Fix a set $\Act$ of local actions and a set $\ff Links$ of links
between agents in $\ff AG$.%
\footnote{We are being a little sloppy here, since we do not
distinguish between an action $a$ and the name for the action that
appears in a program, and similarly for links and the variables in
$X_i$.}
There are five types of basic programs
for agent $i$:
\begin{enumerate}[$\bullet$]
\item ${\cf @} i~{\cf initially~ \psi}$;
\item ${\cf @} i~ {\cf if}~\kind =  k~{\cf then}~ x:= t$,
where $k \in \Act \union {\ff Links}$ and $x \in X_i$;
\item  ${\cf @} i~ \kind= \local(a)~{\cf only~if~} {\cf\phi}$;
\item ${\cf @} i~{\cf if~necessarily}~\phi~{\cf
then~i.o.}~\kind=\local(a) $; and
\item ${ \cf @} i~ {\cf only~events~in}~ L~ {\cf affect}~ x$, where $\ff
L$ is a
list of  kinds in $\Act \union Links$ and $x \in X_i$.
\end{enumerate}
Note that all basic programs for agent $i$ are prefixed by $@i$.

We can form more complicated programs  from simpler programs by {\em
composition}.
We can compose automata associated with different or same agents.
(Note that, since message automata associated with same agent can be
composed in our language, we are not relying on a standard notion of
parallel composition.)
Thus, the set (type) $\ff Pgm$ of programs is the smallest
set that includes the basic programs such that if $\Pg_1$ and
$\Pg_2$ are programs, then so is $\Pg_1 \oplus \Pg_2$.%
\footnote{Here we are deliberately ignoring the difference between
sets and types.}

Readers familiar with UNITY \cite{CM88} will see some obvious
similarities.   In UNITY, a program consists of an initial
condition on a global state, a set of guarded assignment
statements that update this state non-deterministically as if running an
unbounded loop, and a set of allowed actions.  As we said earlier,
communication occurs through reading and writing shared variables
(rather than by message passing, as in Nuprl).  States in Nuprl are also
considerably more expressive than those used in UNITY.

\subsubsection{Semantics}\label{sec:sem}

We give semantics by associating with each program the set of event
structures consistent with it. Intuitively, a set of event
structures is consistent with a distributed message automaton if
each event structure in the set is generated from an execution of the
automaton. The semantics can be defined formally in Nuprl as a
relation between a distributed program $\ff{ Pg}$ and an event
structure $\ff{ es}$.
In this section, we define the consistency relation
for programs and give the intuition behind these programs.

In classical logic, we give meaning to formulas using an interpretation,
that is, an interpretation consists of a domain and an assignment of each
predicate and
function symbol to a predicate and function, respectively, over that
domain.
In the Nuprl setting, we are interested in
\emph{constructive interpretations} $I$, which can be characterized
by a formula $\phi_I$.  We can think of $\phi_I$ as characterizing a
domain $\Val_I$ and the meaning of the function and predicate symbols.
If $I$ is an interpretation with domain $\Val_I$,
an \emph{$I$-local state for $i$}
maps $X_i \union \{\sval_i\}$ to $\Val_I$;
an {\em $I$-global state}
is a tuple of $I$-local states, one for each agent in $\AG$.
Thus, if $s = (s_1, \ldots, s_n)$ is an $I$-global state, then $s_i$ is
$i$'s local state in $s$.
(Note that we previously used $s$ to denote a local state, while here
$s$ denotes a global state.
We will always make it clear whether we are referring to local or
global states.)

For consistency with our later discussion of knowledge-based programs,
we allow the meaning of some predicate and function symbols that appear
in tests in programs to depend on the
global state.  We say that a function or predicate symbol is
\emph{rigid} if it does not depend on the global state.  For example, if
the domain is the natural numbers, we will want to treat $+$, $\times$,
and $<$ as rigid.  However,
having the meaning of a function or predicate depend on the global state
is not quite as strange as it may seem.  For example, we may want to
talk about an array whose values are encoded in agent 1's variables
$x_1$, $x_2$, and $x_3$.  An array is just a function, so the
interpretation of the function may change as the values of $x_1$, $x_2$,
and $x_3$ change.  For each
nonrigid predicate symbol $P\in {\it P}$ and function symbol $f \in
{\it F}$,
we assume that
there is a predicate symbol $P^+$ and function symbol $f^+$ whose arity
is one more than that of $P$ (resp., $f$); the extra argument
is a global state.
We then associate with every formula $\phi$ and term $t$ that appears in
a program a formula $\phi^+$ and term $t^+$ in the language of Nuprl.
We define $\phi^+$ by induction on the structure of $\phi$.  For
example, for an atomic formula such as $P(c)$, if $P$ and $c$ are rigid,
then $(P(c))^+$
is just $P(c)$.  If $P$ and $c$ are both nonrigid, then $(P(c))^+$ is
$P^+(c^+(\bs),\bs)$, where
$\bs$ is interpreted as a
global state.%
\footnote{Since Nuprl is a higher-order language, there is no
problem having a variable  ranging over global states that is
an argument to a predicate.}
We leave to the reader the straightforward task of
defining $\phi^+$ and $t^+$ for atomic formulas and terms.  We then take
$(\phi \land \psi)^+ = \phi^+ \land \psi^+$, $(\neg \phi)^+ = \neg
\phi^+$, and $(\forall x \phi)^+ = \forall x \phi^+$.

An \emph{$I$-valuation} $V$
associates with each non-local variable
(i.e., variable not in $\union_{i \in {\ff AG}} X_i$) a value in
$\Val_I$.
Given an interpretation $I$,
an $I$-global state $s$, and
an $I$-valuation $V$,
\commentout{
 : to determine $I_V(t)(s)$, we replace each
constant ${\cf c}$ other than the $\sval_i$s by the value $I(c)$,
replace  $\sval_i$ by $s_i(\snval_i)$, replace each variable $x$
by $s_i(x)$ if $x\in X_i$ or by $V(x)$ if $x\not \in \union_{i\in
AG}X_i$, and replace each function ${\cf f}$ by the function $I({\cf
f})$. As expected, if $\cf f$ is interpreted as a function of arity
$k$, and $t_1$, $\dots$, $t_k$ are terms, then $I_V({\cf
f}(t_1,\dots,t_k))(s)=I({\cf f})(I_V(t_1)(s),\dots,I_V(t_k)(s))$.
}%
we take $I_V(\phi)(s)$ to be an abbreviation for the formula (expressible
in Nuprl) that says $\phi_I$ together with the conjunction of atomic
formulas of the form $x=V(x)$ for all non-local variables $x$ that
appear in $\phi$, $x=s_i(x)$ for variables $x \in X_i$, $i \in \AG$,
that appear in
$\phi$, and $\bs = s$ implies $\phi^+$.
Thus, $I_V(\phi)(s)$ holds if there is a constructive proof that
the formula that characterizes $I$ together with the (atomic)
formulas that describe $V(x)$ and $s$, and a formula that says that
$\bs$ is represented by $s$, imply $\phi^+$. It is beyond the scope
of this paper
(and not necessary for what we do here)
to discuss constructive proofs in Nuprl; details can be found in
\cite{book}.
However, it
is worth noting that,
for a first-order formula $\phi$, if $I_V(\phi)(s)$ holds, then
$\phi^+$
is true in state $s$ with respect to the semantics of classical logic
in $I$.
The converse is not necessarily true.  Roughly speaking,
$I_V(\phi)(s)$ holds if there is evidence for the truth of $\phi^+$ in
state $s$ (given valuation $V$).  We may have evidence for neither
$\phi^+$ nor $\neg \phi^+$.

A formula $\phi$ is an \emph{$i$-formula in interpretation $I$} if its
meaning in $I$ depends only in $i$'s local state;
that is, for all global states
$s$ and $s'$ such that $s_i=s'_i$, $I_V(\varphi)(s)$ holds iff
$I_V(\varphi)(s')$ does.
Similarly, $t$ is an \emph{$i$-term in $I$} if $x=t$ is an $i$-formula
in $I$,
for $x$ a non-local variable.
It is easy to see that $\phi$ is an $i$-formula in all interpretations
$I$ if all the predicate and function symbols in $\phi$ are rigid, and
$\phi$ does not mention variables in $X_j$ for $j \ne i$ and does not
mention the constant symbol $\sval_j$ for $j \ne i$.
Intuitively, this is because if we have a constructive proof that
$\phi$ holds in $s$ with respect to valuation $V$, and
$\phi$ is an $i$-formula, then all references to local states of
agents other than $i$ can be safely discarded from the argument to
construct a proof for $\phi$ based solely on $s_i$.
If $\phi$ is an $i$-formula, then we sometimes abuse notation
and write $I_V(\varphi)(s_i)$ rather than $I_V(\varphi)(s)$. Note
that the valuation $V$ is not needed for interpreting formulas
whose free variables are  all local; in particular, $V$ is not
needed to interpret $i$-formulas. For the rest of this
paper, if the valuation is not needed, we do not mention it, and
simply write
$I(\varphi)$.
Given a formula $\phi$ and term $t$, we
can easily define Nuprl formulas \emph{i-formula}($\phi$,$I$) and
\emph{i-term}($t$,$I$) that are constructively provable if $\phi$ is an
$i$-formula in $I$ (resp., $t$ is an $i$-term in $I$).

We define a predicate $\Consistent_I$ on programs and event
structures such that, intuitively, $\Consistent_I$ $(\Pg,\es)$ holds
if the event structure $\es$ is consistent with program $\Pg$, given
interpretation $I$. We start with basic programs. The basic program
${\cf @} {\ff i}~{\cf initially~\psi}$ is an {\em initialization}
program, which is intended to hold in an event structure $\es$ if
$\psi$ is an $i$-formula and
$i$'s initial
local
state satisfies $\psi$. Thus,
$$
\begin{array}{l}
\Consistent_I( {\cf @} {\ff i}~ {\cf initially~\psi}, \es)
\eqdef i\mbox{-\emph{formula}}(\psi,I) \land I(\psi)(\initstate_i).
\end{array}
$$
(This notation implicitly assumes that
$\initstate_i$ is as specified by $\es$, according to
Definition~\ref{def:main}. For simplicity, we have opted for this
notation instead of
${\es}.{\initstate_i}$.)
We call a basic program of the form ${\cf @} i~ {\cf if}~\kind =
k~{\cf then}~ x:= t$ an {\em effect} program.  It says that,
if $t$ is an $i$-term, then
the
effect of an event $\ff e$ of kind $k$ is to set $\ff x$ to $t$. We
define
$$
\begin{array}{l}
\Consistent_I( {\cf @} i~ {\cf if}~\kind =  k~ {\cf then}~ x:= t,
\es)
\eqdef\\
\quad i\mbox{-\emph{term}}(t,I) \land
\ff{\forall e@i\in es.~(\kind(e)=k \Rightarrow
 (\state ~\esafter~e)(x) =I(t)(\state ~\esbefore~e)),}
\end{array}
$$
where we write $\ff \forall e@i\in es. ~\phi$ as an abbreviation for
$\forall e \in es. \agent(e) = i \rimp \phi$.
The
notation above implicitly assumes that $\esbefore$ and
$\esafter$ are as specified by $\es$.
Again, this expression is an abbreviation for a formula expressible
in Nuprl whose intended meaning should be clear; $\Consistent_I(
{\cf @} i~ {\cf if}~\kind =  k~ {\cf then}~ x:= t, \es)$ holds if
there is a constructive proof of the formula.

We can use a program of this type to describe a message sent on a
link $l$.  For example,
$${\cf @}i~ \kind=\local(a)~{\cf then~msg}(l) {\cf \, := \,
f}(\sval_i)$$ says that for all events $e$, $f(v)$ is sent on link
$l$ if the kind of $e$ is $a$, the
local
state of agent $i$ before $e$ is $s_i$, and $v = s_i(\snval_i)$.

The third type of program, ${\cf @} i~ \kind= \local(a)~{\cf
only~if~} {\cf\phi}$, is called a {\em precondition} program. It
says that an event of kind $a$ can occur only if the precondition
$\varphi$
(which must be an $i$-formula)
is satisfied:
$$
\begin{array}{l}
\Consistent_I ({\cf @}i ~\kind=\local(a)~{\cf only~if}~\varphi, \es)
\eqdef\\
\quad i\mbox{-\emph{formula}}(\phi,I) \, \land
\forall e@i\in \es.~ (\kind(e)=\local(a) \Rightarrow
I(\varphi)(\state~\esbefore~e)).
\end{array}
$$
Note that we allow conditions of the form $\kind(e) = \local(a)$
here, not the more general condition of the form $\kind(e) = k$
allowed in effect programs.  We  do not allow conditions of the form
 $\kind(e) = \rcv(l)$ because we assume that receive events are not
under the control of the agent.

Standard formalizations of input-output automata (see \cite{LT})
typically assume that executions satisfy some fairness constraints.
We assume here only a weak fairness constraint that is captured by
the basic program ${\cf @} i~{\cf if~necessarily}~\phi~{\cf
then~i.o.}~\kind=\local(a)$,
 which we call a \emph{fairness program}.
Intuitively, it says that if $\phi$ holds from some point on, then
an event with kind $\local(a)$ will eventually occur.  For an event
sequence with only finitely many states associated with $i$, we take
$\phi$ to hold ``from some point on'' if $\phi$ holds at the last
state.  In particular, this means that the program cannot be
consistent with an event sequence for which there are only finitely
many events associated with $i$ if $\phi$ holds of the last
state associated with $i$.
Define
$$
\begin{array}{l}
\Consistent_I({\cf @} i~{\cf if~necessarily}~\phi~{\cf
then~i.o.}~\kind=\local(a), \es) \eqdef\\
\quad i\mbox{-\emph{formula}}(\phi,I) \, \land\\
\quad    [((\exists e@i \in \es) \land \forall e@i\in \es. ~ \exists
e'\essucceqi e.~ I(\neg \phi)     (\state~\after~e') \vee (\kind(e')=\local(a)) ) \\
\quad       \vee\,
(\neg (\exists e@i\in \es) \land I(\neg
\phi)({\initstate}_i)) ].
\end{array}
$$
\noindent The last type of basic program, ${\cf @i~only~events~in~ L~
  affect~ }x$, is called a {\em frame program}.  It ensures that only
events of kinds listed in $\ff L$ can cause changes in the value of
variable $\ff x$. The precise semantics depends on whether $x$ has the
form $\msg(l)$. If $x$ does not have the form $\msg(l)$, then
$$
\begin{array}{l}
{\ff \Consistent_I(} {\cf @} {\ff i}~{\cf only~ events~in~L~ affect~ }x  {\ff
, es)\eqdef}\\
\quad    \ff{\forall e@i\in es.~ ((x~\esafter~e)\neq
(x~\esbefore~e)}{\ff \rimp (\kind(e)\in L)). }
\end{array}
$$

\noindent If $x$ has the form $\msg(l)$, then we must have $\source(l) = i$.
Recall that sending a message $m$ on $\ff l$ is formalized by
setting the value of $\ff \msg(l)$ to $m$.  We assume that messages
are never null (i.e., $m \ne \bot$).  No messages are sent during
event $e$ if $\msg(l) \after~ e= \bot$. If $x$ has the form
$\msg(l)$, then
$$
\begin{array}{l}
{\ff \Consistent_I(} {\cf @} {\ff i}~{\cf only~events~in~ L~ affect~ msg}(l)
{\ff
, es)\eqdef}\\
\quad    \ff{\forall e@i\in es.~ ((\msg(l)~\esafter~ e \neq
\bot)}{\ff \rimp (\kind(e)\in L))}.
\end{array}
$$

\noindent Finally, an event structure $\ff es$ is said to be consistent with a
distributed program $\ff Pg$ that is not basic if $\ff es$ is
consistent with each of the basic programs that form $\ff Pg$:
$$
\begin{array}{l}
{\ff \Consistent_I(Pg_1 \oplus Pg_2, es)\eqdef
\Consistent_I(Pg_1, es)\wedge \Consistent_I(Pg_2, es)}.
\end{array}
$$

\dfn\label{def:sem} Given an interpretation $I$, the semantics of a
program $\Pg$ is the set of event structures consistent with $\Pg$
under interpretation $I$. We denote by $S_I$ this semantics of
programs: $\ff{S_I(\Pg)=\lbrace es \: |\: }$ $\ff{\Consistent_I(\Pg,
es)\rbrace}.$ We write $\Pg \psatI X$ if $\Pg$ satisfies $X$ with
respect to interpretation $I$; that is,
if $\ff{X}(\es)$ is true for all $\es\in \ff{S_I}(Pg))$.
\edfn Note
that $\ff{S_I(Pg_1\oplus Pg_2)=S_I(Pg_1)\cap S_I(Pg_2)}.$
Since the $\ff{ \Consistent_I}$ predicate is
definable in Nuprl, we can formally reason  in Nuprl about
the semantics of programs.

A specification is a predicate on event structures.
Since our main goal is to derive from a proof that a specification
$\ff X$ is satisfiable by a program that satisfies $\ff X$, we want to
rule out
the trivial case where the derived program $\Pg$ has no executions,
so that it vacuously satisfies the specification $\ff X$.

\dfn\label{def:realiz} Program $\Pg$ is {\em \pgrealizable\ (with
respect to interpretation $I$)} if
$S_I(\Pg)\neq \emptyset$.  The specification  $\ff X$ is {\em
realizable (with respect to interpretation $I$)} if it is not
vacuously satisfied, that is, if $\ff{\exists Pg. (Pg\psatI X\wedge
} {\ff S_I(Pg)\neq \emptyset)}$. $\Pg$ \emph{realizes} $X$ (with
respect to $I$) if $\Pg \psatI X$ and $\Pg$  is \pgrealizable\ (with
respect to $I$). \edfn

Thus, a specification is realizable if there exists a consistent
program that satisfies it, and, given an interpretation $I$, a
program is realizable if there exists an event structure consistent
with it (with respect to $I$). Since we reason constructively, this
means that a program is realizable if we can {\em construct} an
event structure consistent with the program. This requires not only
constructing sequences of events, one for each agent, but all the
other components of the event structure as specified in
Definition~\ref{def:ev structure},
such as $\mathit{AG}$ and $\Act$.

\commentout{
With the exception of the initialization clause, all basic clauses
are trivially {\pgrealizable}: we simply take  $\es$ to be an event
structure with no events; $\es$ is consistent with such basic
clauses since their semantics in defined as a universal
quantification over events associated with an agent. (We remark that
this is different from the original approach of
Constable and Bickford~\cite{CB08}. Following the definition of fair
tasks in
message-automata literature~\cite{LT}, in \cite{CB08} the semantics
of fair clauses ${\cf @}i~{\cf i.o.}~\kind=\local(a)~{\cf
if~i.o.}~\varphi$ rules out finite event structures $\es$ such that
$\varphi$ is enabled after the last event associated with $i$ in
$\es$. Under this semantics, an event structure with no events does
not satisfy the fair clause; furthermore, no event structure with
finitely many events associated with agent $i$ satisfies the fair
clause, if the condition $\varphi$ is infinitely often enabled.
Nevertheless, provided $\varphi$ satisfies the principle of excluded
middle, Bickford and Constable were able to carry out in Nuprl the
construction of an event structure consistent with the fair clause.
The details of the construction are beyond the scope of this paper.
Furthermore, for simplicity, the constraint on ruling out finite
event structures has been dropped in the revised version of the
technical report~\cite{EvSys05}.) }
All basic programs other than initialization and fairness programs
are vacuously satisfied
(with respect to every interpretation $I$)
by the empty event structure $\es$
consisting of no events.  The empty event structure is consistent
with these basic programs because their semantics in defined in
terms of a universal quantification over events associated with an
agent.
It is not hard to see that an
initialization program ${\cf @} {\ff i}~ {\cf initially~\psi}$ is
\pgrealizable\ with respect to interpretation $I$ if and only if
$\psi$ is satisfiable in $I$; i.e., there is some global state $s$ such
that $I(\psi)(s_i)$ holds.
For if $\es$ is an
event structure with $\initstate_i = s_i$, then
clearly $\es$ realizes ${\cf @} {\ff i}~ {\cf initially~\psi}$.

Fair programs are realizable with respect to interpretations $I$
where the precondition $\phi$ satisfies the principle of excluded
middle (that is, $\phi_I \rimp \Dec(\phi^+)$ is provable in Nuprl),
although they are not necessarily realized by a finite event structure.
To see this, note that
if $\phi$ satisfies the principle of excluded middle in
$I$, then either there is an $I$-local
state $s^*_i$ for agent $i$ such that
$I(\neg \phi)(s^*_i)$ holds,
or $I(\phi)(s_i)$ holds for all $I$-local states $s_i$ for $i$.
In the former case,
consider an empty event structure $\es$ with domain $\Val_I$ and
$\initstate_i = s^*_i$; it is easy to see that $\es$ is consistent
with ${\cf @} i~{\cf if~necessarily}~\phi~{\cf then~} $ ${\cf
i.o.}~\kind=\local(a)$. Otherwise, let $\Act= \lbrace a \rbrace$.
Let $\es$ be an event structure where $\Act$ is the set of local
actions, $\Val_I$ is the set of values,
the
sequence of events associated with agent $i$ in $\es$ is infinite, and
all events associated with agent $i$ have kind $\local(a)$.  Again, it is easy
to see that $\es$ is consistent with ${\cf @} i~{\cf if~necessarily}~\phi~{\cf
then~i.o.}~\kind=\local(a)$.

If $\phi$ does not satisfy the principle of excluded
middle in $I$, then $${\cf @} i~{\cf if~necessarily}~\phi~{\cf
then~}{\cf i.o.}~\kind=\local(a)$$ may not be realizable with
respect to $I$.
This
would be the case
if, for example,
neither $I(\phi)(s_i)$ nor $I(\neg \phi)(s_i)$ holds for any local
state $s_i$.

Note that two initialization programs may each be \pgrealizable\
although their composition is not.
For example, if both $\psi$ and $\neg \psi$ are satisfiable
$i$-formulas,
then
each of ${\cf @} {\ff i}~ {\cf initially~\psi}$ and ${\cf @} {\ff
i}~ {\cf initially~\neg \psi}$
is consistent, although their composition is not. Nevertheless, all
programs synthesized in this paper can be easily proven
\pgrealizable.

\subsubsection{Axioms}\label{sec:axioms}
Constable and Bickford \cite{CB08}  derived from the formal
semantics of distributed message automata some Nuprl axioms that
turn out to be useful for proving the satisfiability of a
specification. We now present (a slight modification of) their
axioms.
The axioms have the form $\Pg \psatI X$, where $\Pg$ is a program
and $X$ is a specification, that is, a predicate on event
structures; the axiom is sound if all event structures $\es$ consistent
with program $\Pg$ under interpretation $I$ satisfy the specification
$X$.
We write $\psatI$ to make clear that the program semantics is given
with respect to an interpretation $I$.
There is an axiom for each type of basic program other than frame
programs, two axioms for frame programs (corresponding to the two
cases in the semantic definition of frame programs), together with
an axiom characterizing composition and a refinement axiom.
\begin{enumerate}[\hbox to8 pt{\hfill}]
\item\noindent{\hskip-12 pt\bf Ax\mbox{-}init::}\
$${\cf @} {\ff i} ~{\cf   initially ~  \psi} \ff{ \, \psatI }
\ff{\lambda \es. ~i\mbox{-\emph{formula}}(\psi,I) \, \land
I(\psi)(\initstate_i)}.$$
(Note that the right-hand side of $\psat$ is a specification; given
an event structure $\ff es$,
it is true if $i\mbox{-\emph{formula}}(\psi,I) \, \land
I(\psi)(\initstate_i)$ holds in event structure $\es$.)

\item\noindent{\hskip-12 pt\bf Ax\mbox{-}cause::}\
$$\begin{array}{ll}
                 {\cf @}i ~{\cf if}~ \kind=k~{\cf then}~x {\cf  :=} t
                         \ff{\, \psatI } &
\ff{\lambda es.~i\mbox{-\emph{term}}(t,I) \, \land
 \forall
e@i\in \es.~(\kind(e)=k  \Rightarrow } \\
                                        &
                   \ff{(\state ~\esafter~e)(x) =I(t)(\state~\esbefore~e)).}
\end{array}
$$
\item\noindent{\hskip-12 pt\bf Ax\mbox{-}if::}\
$$
\begin{array}{l}
 {\cf @}i ~\kind=\local(a)~ {\cf only~if}~\varphi \ff{\, \psatI} \\
 \quad \begin{array}{ll}
   {\ff \lambda es.} & {\ff i\mbox{-\emph{formula}}(\phi,I) \, \land } \\
                     & {\ff \forall e@i\in es.~(\kind(e)=\local(a)} \Rightarrow 
                 {\ff I(\phi)(\state~\esbefore~e))}.
        \end{array}
\end{array}
$$
\commentout{
\item\noindent{\hskip-12 pt\bf Ax\mbox{-}StrongFair:}\
$$
\begin{array}{l}
{\cf @}i~{\cf i.o.}~\kind=\local(a)~{\cf if~i.o.}~\varphi \ff{\, \psatI} \\
  \quad  \begin{array}{ll}
         \ff{\lambda es.} & \begin{array}{ll}
                            \forall e@i\in es. & ({\ff \exists e' \essucci e.~(\kind(e')=\local(a))}) ~\vee
                            \end{array} \\
                          & \begin{array}{ll}
                                               & ({\ff \exists e'' \essucceqi
  e. ~\forall e' \essucci e''.~
                                                 I(\neg
  \phi)(\state~\esafter~e')})
                            \end{array} \\
                          & ~ \wedge (\mathit{finite}_i(\es) \rimp I(\neg \phi)(\mathit{laststate}_i(\es))).
         \end{array}\\
\end{array}
$$
}
\item\noindent{\hskip-12 pt\bf Ax\mbox{-}fair::}\
$$
\begin{array}{l}
{\cf @} i~{\cf if~necessarily}~\phi~{\cf
then~i.o.}~\kind=\local(a) \ff{\, \psatI} \\
  \quad  \begin{array}{ll}
          \ff{\lambda es.} & {i\mbox{-\emph{formula}}(\phi,I)} \land\\
             & [ ( (\exists e@i \in \es) \land \\
             & \quad \forall e@i\in \es. ~ \exists e'\essucceqi
e.~I(\neg \phi) (\state~\after~e') \vee (\kind(e')=\local(a)) )~
\\
&
\vee  (\neg (\exists e@i\in \es) \land I(\neg
\phi)({\initstate}_i))].
         \end{array}\\
\end{array}
$$
\item\noindent{\hskip-12 pt\bf Ax\mbox{-}affect::}\ 
$$
 \begin{array}{l}
      {\cf @} {\ff i}~ {\cf only~events~in~ L~ affect ~ } x \ff{\, \psatI }\\
                           \quad \ff{\lambda es.~\forall e@i\in
      es. ~(x~\esafter~e\neq x~\esbefore~e)}
    {\ff \rimp (\kind(e)\in L). }
                          \end{array}$$
\item\noindent{\hskip-12 pt\bf Ax\mbox{-}sends::}\
$$
 \begin{array}{l}
   {\cf @} {\ff i}~ {\cf only~events~in~ L~ affect ~ msg}
                         \ff{(l)\, \psatI } \\
                           \quad \ff{\lambda es.~\forall e@i\in es.~
                         (\msg(l)~\esafter~e \neq \bot)}
   {\ff \rimp (\kind(e)\in L). }
                          \end{array}$$

\end{enumerate}
\begin{enumerate}[\hbox to8 pt{\hfill}]
\item\noindent{\hskip-12 pt\bf Ax-$\oplus$::}\
 $\ff{(Pg_1\psatI P)\wedge (Pg_2\psatI Q)
\Rightarrow  (Pg_1\oplus Pg_2 \, \psatI \, P\wedge Q).}  $
\item\noindent{\hskip-12 pt\bf Ax\mbox{-}ref::}\
 $\ff{(Pg \, \psatI \, P)\wedge (P\Rightarrow
Q) \Rightarrow (Pg\, \psatI \, Q). } $
\end{enumerate}

\begin{lem}\label{lem:ax sound} Axioms ${\bf Ax\mbox{-}init}$, ${\bf
Ax\mbox{-}cause}$,
${\bf Ax\mbox{-}if}$,
${\bf Ax\mbox{-}fair}$,
${\bf Ax\mbox{-}affect}$, ${\bf Ax\mbox{-}sends}$,
${\bf Ax\mbox{-}\oplus}$, and ${\bf Ax\mbox{-}ref}$ hold for all
interpretations $I$. \end{lem} 

\proof This is immediate from
Definitions~\ref{def:main} and~\ref{def:sem}, and the definition of
$\ff\Consistent_I$. \qed

\subsubsection{A general scheme for program synthesis}\label{sec:schemes}
Recall that,  given a specification $\ff \varphi$ and an
interpretation $I$, the goal is to prove that $\ff \varphi$ is
satisfiable with respect to $I$, that is, to show that $\ff \exists
Pg.~ (Pg \psatI \varphi)$ holds. We now provide a general scheme for
doing this. Consider the following scheme, which we call $\it{GS}$:
\begin{enumerate}[(1)]
\item Find specifications $\ff \varphi_1$, $\ff \varphi_2$, $\dots$, $\ff \varphi_n$ such that
$\ff \forall es.~( \varphi_1(es) \wedge \varphi_2(es) \wedge \dots
\wedge \varphi_n(es) \Rightarrow \varphi(es)) $ is true under
interpretation $I$.
\item Find programs $\ff Pg_1$, $\ff Pg_2$, $\dots$, $\ff Pg_n$ such that
$\ff Pg_i \psatI \varphi_i$ holds for all $\ff i \in \{1, \ldots
n\}$.
\item Conclude that $\Pg \psatI \varphi$, where $\Pg=\Pg_1 \oplus \Pg_2
\oplus \dots \oplus \Pg_n$.
\end{enumerate}\medskip

Step 1 of $\it{GS}$ is proved using the rules and axioms encoded in
the Nuprl system; Step 2 is proved using the axioms given in
Section~\ref{sec:axioms}. It is easy to see that $\it{GS}$ is {\em
sound} in the sense that, if we can show using $\it{GS}$ that $\Pg$
satisfies $\phi$, then $\Pg$ does indeed satisfy $\phi$. We
formalize this in the following proposition.

\pro\label{pro:scheme} Scheme $\it{GS}$ is sound. \epro

\subsection{Example}\label{example: spec}
As an example of a specification that we use later, consider the
run-based specification
$\ff{ Fair_I(\varphi,t,l)}$, where
$i \ne j$,
$\ff{ l}$ is a link with $\ff \source(l)=i$ and $\ff
\dest(l)=j$, $\varphi$ is an $i$-formula, and $\ff t$ is an $i$-term.
$\ff{Fair_I(\varphi,t,l)}$ is a conjunction of a safety condition
and a liveness condition. The safety condition asserts
that if a message is received on link $\ff{ l}$, then it is the term
$\ff{ t}$ interpreted with respect to the
local
state of the sender, and that $\varphi$, evaluated with
respect to the
local
state of the sender, holds.
(More precisely, $\phi$ holds when evaluated with respect to the state
of agent $i$ before $e$ occurs, that is,
in $\state ~\before~e$.)
The liveness condition says that, if
(there is a constructive proof that)
condition $\phi$ is enabled
from some point on in an infinite event sequence, then eventually a
message sent on $\ff l$ is delivered. (Thus, the specification
imposes a weak fairness requirement.)
We define $\ff Fair_I(\varphi,t,l)$ as follows:
$$
\begin{array}{l}
{\ff Fair_I(\varphi,t,l) \eqdef}  {\ff \lambda
es. ~i\mbox{-\emph{formula}}(\varphi,I) \land
i\mbox{-\emph{term}}(t,I) \land }\\
\    {\ff (\forall e'\in es.~(\kind(e')=\rcv(l)\Rightarrow } \\
\quad         {\ff I(\varphi)(\state
                    ~\before~\send(e'))}
                         \wedge  {\val(e')=I({\ff
                    t})(\state~\before~\send(e'))})
                     ~ \wedge\\
\commentout{
                      \begin{array}{ll}
                      \wedge ({\ff \forall e@i\in es.}
                                & ({\ff \exists e'.~
                                   (\send(e') \essucci  e \land \kind(e')=\rcv(l)) })
                                   \vee \\
                                &  (\exists e'' \essucceqi e.~\forall e'
                    \essucci e''.
~ I(\neg \varphi)(\state~\after~e')).
                      \end{array} \\
}
(   ( \exists e@i \in \es \land \\
\quad \forall e@i \in \es. ~ \exists e'
\essucceqi e.~
                                        I(\neg \phi)(\state  ~\after~e'))
     \, \vee \,
    (\neg (\exists e@i\in \es) \land  I( \neg \phi)({\initstate}_i)) \\
\quad \vee \,
   (\exists e@i \in \es \land  \forall e@i \in \es. ~\exists e' \succeq_i e .~\kind(e')=\rcv(l)
                                     \wedge \send(e') \essucceqi e) ).
\end{array}
$$
We are interested in this fairness specification only in settings
where communication satisfies a (strong) fairness requirement: if
infinitely often an agent sends a message on a link $l$, then
infinitely often some message is delivered on $l$. We formalize this
assumption using the following specification:
\commentout{ We are interested in this fairness specification only
in settings
where communication satisfies a (strong) fairness requirement: if an
agent sends the same message on link $l$ infinitely often, then the
message is delivered infinitely often. We formalize this assumption
using the following specification: }
\commentout{
$$
\begin{array}{ll}
\FairS(l) \eqdef \lambda \ff{es}. &(\forall e@i \in \ff{es}.
~\exists e' \essucci e.~   (\msg(l)~\after~e'\neq \bot))
\Rightarrow \\
&\ff{  (\forall e@i \in es. ~\exists e'.~ (\kind(e')=\rcv(l)\wedge
\send(e') \essucci e))}.
\end{array}
$$
}
$$
\begin{array}{ll}
\FairS(l) \eqdef \lambda \ff{es}. & (\forall e@i \in \ff{es}.
~\exists e' \essucci e.~
\msg(l)~\after~e' \neq \bot )\\
&  \begin{array}{ll}
   \rimp \ff{  (\forall e@i \in es. ~\exists e'.} &
                \ff{\kind(e')=\rcv(l)\wedge \send(e') \essucci e)}.
    \end{array}
\end{array}
$$
We explain below
why
we need communication to satisfy strong fairness rather
than weak fairness (which would require only that if
a message is sent infinitely often, then a message is eventually
delivered).

In this section, we show that, assuming that the communication on
link $l$ satisfies a strong fairness requirement, the specification
above is satisfiable, and that a program that satisfies it can be
formulated in our language. Furthermore, we show that there are simple
conditions on the formulas involved in this program that ensure the
existence of at least one execution of the program. Though the
specification above and the program that satisfies it refer to a
single agent, we show that it is not difficult to extend our results
to a system with
many agents.

For an arbitrary action $a$, let  $\ff{ Fair\mbox{-}Pg(} \varphi,
{\ff t ,l,a)}$ be the following program for agent $\ff{ i}$:
$$\begin{array}{l}
{\cf @}i~\kind=\local(a)~{\cf only ~if}~{\cf \varphi} ~\oplus\\
{\cf @}i~{\cf if }~ \kind=\local(a)~{\cf then}~ {\cf msg}(l){\cf :=} {\ff t} ~\oplus\\
{\cf @}i~{\cf only~ events~ in ~ [ } a{\cf ] ~ affect ~msg(}l{\cf )} ~\oplus\\
{\cf @}i~{\cf if~necessarily}~\phi~{\cf then~i.o.}~\kind=\local(a).\\
\end{array}
$$
The first basic program says that $i$ takes action $a$ only if
$\varphi$ holds. The second basic program says that the effect of
agent $i$ taking action $a$ is for ${\ff t}$ to be sent on link $l$;
in other words, $a$ is $i$'s action of
sending ${\ff t}$ to agent $j$. The third program ensures that only
action $a$ has the effect of sending a message to agent $j$. With
this program, if agent $j$ (the receiver) receives a message from
agent $i$ (the sender), then it must be the case that
the value of the message is ${\ff t}$ and that $\varphi$ was true
with respect to $i$'s  local state when it sent the message to $j$.
The last basic program ensures
that if
\commentout{ $\varphi$ holds infinitely often, then $i$ takes action
$a$ infinitely often, that is, sends ${\ff t}$ to $j$.
\commentout{ with respect to $i$'s state
and some value, }
then eventually $j$ will receive $i$'s message. }
$\phi$ holds from some point on in an infinite event sequence, then
eventually an event of kind $a$ holds; thus, $i$ must send the
message $t$ infinitely often.
The fairness requirement on communication ensures that if an event
of kind $a$ where $i$ sends $t$ occurs infinitely often, then $t$ is
received infinitely often.

\begin{lem}\label{lem:ex fair} For all actions $\ff a$, $\ff
{Fair\mbox{-}Pg(} \varphi{\ff , t ,l,a)} $ satisfies $$\ff \lambda
\es. \FairS(l)(\es) \rimp
Fair_I(\varphi,t,l)(\es)$$ with
respect to all interpretations $I$ such  that $\phi$ is an
$i$-formula and $t$ is an $i$-term in $I$.\end{lem}

\proof We present the key points of the proof here, omitting some
details for ease of exposition. We follow the scheme $\it{GS}$.
We assume that $i\mbox{-\emph{formula}}(\phi,I)$ and
$i\mbox{-\emph{term}}(t,I)$ both hold.

{\bf Step 1.} For each event structure $\ff es$, $\ff
Fair_I(\varphi,t,l)(es)$ is equivalent to a conjunction of three
formulas:
$$
\begin{array}{lll}
{\ff  \phi_1(es) }:
  &{\ff \forall e'\in es. ~(\kind(e')=\rcv(l)) \Rightarrow
    I(\varphi)(state~\before ~\send(e'))}\\
{\ff  \phi_2(es) }:
   & {\ff \forall e'\in es.~ (\kind(e')=\rcv(l)) \Rightarrow
   \val(e')=I(t)(\state ~\before ~\send(e'))} \\
{\ff  \phi_3(es)}: &
     ( \exists e@i \in \es \land  \forall e@i\in es. ~\exists e' \essucceqi e.~
     I( \neg \phi)(\state ~\after~e')) \, \vee \\
& (\neg (\exists e@i\in \es) \land I(\neg \phi)({\initstate}_i) \, \vee\\
        & (\exists e@i \in \es \land \forall e@i\in \es. ~\exists e'. \kind(e')=\rcv(l) \wedge
      \send(e') \essucci e ).
\end{array}
$$
We want to find formulas $\psi_1(es), \ldots, \psi_4(es)$ that
follow from the four basic programs that make up $\ff{
Fair\mbox{-}Pg(}\varphi {\ff , t, l,a)}$ and together imply
$\phi_1(es) \land \phi_2(es) \land \phi_3(es)$. It will simplify
matters to reason directly about the events where a message is sent
on link $l$. We thus assume that, for all events $e$, agent $i$
sends a message on link $l$ during event $e$ iff $\kind(e) = {\ff
local}(a)$.
This assumption is expressed by:
$$   {\ff \psi_1(\es) \eqdef \forall e@i\in \es.~ (\msg(l)~ \after ~e \neq \bot)
          \rimp (\kind(e)=\local(a))}.$$
It is easy to check that $(\psi_1(\es) \land \psi_2(\es)) \rimp
\phi_1(\es))$ is true, where $\psi_2(\es)$ is
 $${\ff \forall e@i\in \es}.~ (\kind(e)=\local(a)) \Rightarrow
I(\varphi)(\state~ \before~ e).$$ Similarly, using the axiom of
event structures given in Section~\ref{sec:eventsystems} that says
that the value of a receive event $\ff e$ on $\ff l$ is the value of
$\ff \msg(l)$ after $\ff \send(e)$, it is easy to check that
$(\psi_1(\es) \land \psi_3(\es)) \rimp \phi_2(\es))$ is true, where
$\psi_3(\es)$ is
$${\ff \forall e@i\in \es. }
     {\ff (\kind(e)=\local(a)) \Rightarrow }
   {\ff \msg(l)~ \after ~e=I(t)(\state~ \before~ e)}.
$$
We can show that $(\psi_3(\es) \land \psi_4(\es) \land \FairS(l))
\rimp \phi_3(\es)$ is true, where $\psi_4$ is
\commentout{
$$
\begin{array}{l}
 \begin{array}{ll}
 {\ff \forall e@i\in \es.} &  (\exists e'\essucci e.~ \kind(e')=\local(a))
                              \vee \\
                           &  (\exists e''\essucceqi e.~\forall
                                  e'\essucci e. ~
                ~ I(\neg \varphi)(\state~\esafter~e')).
 \end{array}\\
\end{array}
$$
}
$$
\begin{array}{l}
(\exists e@i \land \forall e@i\in es.~ \exists e'\essucceqi e.~
I( \neg \phi)(\state~\after~e')) \, \vee \\
   (\neg (\exists e@i \in es) \land   I(\neg
   \phi)({\initstate}_i)) \, \vee\\
(\exists e@i \in \es \land  \forall e@i\in es.~ \exists e'\essucceqi
e.~\kind(e')=\local(a)).
\end{array}
$$

It follows that $$ \ff{ (\forall \es. (\psi_1(\es) \land \psi_2(\es)
\land \psi_3(\es) \land \psi_4(\es)) \rimp (\FairS(l)(\es) \rimp
Fair_I(\varphi,t,l)(\es)))}.$$

{\bf Step 2.}
By  $\bf{Ax\mbox{-}sends}$, $${\cf @} {\ff i} ~{\cf only ~events ~in~[}
{\ff a}
{\cf ]~affect~msg(} {\ff l} {\cf )} {\ff
\psatI \psi_1}.$$
By $\bf Ax\mbox{-}if$,
$${\cf @}i~\kind=\local(a)~{\cf only~if~\varphi} \psatI \psi_2.$$
By $\bf{Ax\mbox{-}cause}$,
$${\cf @}i ~{\cf if}~\kind=\local(a)~{\cf then}~{\cf msg}(l) {\cf :=} t ~
{\psatI \psi_3};$$
and by $\bf Ax\mbox{-}fair$
$${\cf @}i~{\cf if~necessarily ~}\phi~{\cf then~i.o.}~\kind=\local(a) \psatI \psi_4.$$

\commentout{
 {\bf Step 3.} It is straightforward to check that the
four
basic programs whose composition is
$\ff {Fair\mbox{-}Pg(} \varphi{\ff , t, l,a)} $
are pairwise compatible: since all
four programs are associated with the same agent,
it is easy to see that the only
constraints mentioned by Definition~\ref{def:compat} that we have to
check
are
whether the kind $\ff a$ is a member of the list of kinds mentioned
in the frame program, which
is clearly true,
and that the precondition and fairness requirement programs refer to
the same precondition, which is again true.

{\bf Step 4.}
If $\ff \forall s. \Dec(\exists v. I(\varphi)(v,s))$ holds,
then by Definition~\ref{def:feasible}, all
four
basic programs
whose composition is $\ff{ Fair\mbox{-}Pg(} \varphi{\ff  , t ,l,a)}$
are feasible
with respect to $I$.
}%

\noindent By the soundness of ${\it GS}$ (Proposition~\ref{pro:scheme}),
$\ff{Fair\mbox{-}Pg(\varphi, t, l,a)}$ satisfies
$\lambda \es. \ff{\FairS(l)}(\es) \rimp \ff
Fair_I(\varphi,t,l)(\es)$
with respect to $I$. \qed

\commentout{ It is not difficult to notice that ${\ff
Fair\mbox{-}Pg(}\varphi{\ff ,t ,l,a)}$ is {\pgrealizable}, and thus
$\lambda \ff{\FairS(l)}(\es) \rimp \ff Fair(I(\varphi),I(t),l)(\es)$
is realizable with respect to $\I$.}

\begin{lem}\label{lem: fair pg consistent}
For all interpretations $I$ such that $\phi$ is an $i$-formula and
$t$ is an $i$-term in $I$, if
$\phi$ satisfies the principle of
excluded middle with respect to $I$, then ${\ff
Fair\mbox{-}Pg(}\varphi{\ff,t,l,a)}$ is {\pgrealizable} with respect
to $I$. \end{lem}

\proof
This argument is almost identical to that showing that fair programs are
realizable with respect to interpretations where the precondition
satisfies the principle of excluded middle.
Since $\phi$ satisfies the principle of excluded middle with respect
to $I$, either
there exists  an $I$-local state
$s^*_i$ for agent $i$ such that
$I(\neg \phi)(s^*_i)$ holds,
or $I(\phi)(s_i)$ holds for all
$I$-local states $s_i$ for $i$.
In the former case, let $\es$ be an empty event
structure such that
$i, j \in \AG$, $l \in \Links$,
$a \in Act$, and $\initstate_i = s^*_i$.  In the latter case,
choose $\es$ with
$\AG$ and $\Links$ as above, let $\Act = \{a,b\}$, and
where $i$ and $j$ alternate sending and receiving the
message $t$ on link $l$, where these events have kind $a$ and
$b$, respectively.
\qed

\begin{corollary}\label{cor: fair spec realizable}
For all interpretations $I$ such that if $\phi$ is an $i$-formula
and $t$ is an $i$-term in $I$,  if
$\phi$ satisfies the principle of excluded middle with respect to
$I$, then the
specification $\ff Fair_I(\varphi,t,l)$ is realizable with
respect to $I$. \end{corollary}
\proof This is immediate from Lemmas~\ref{lem:ex fair} and \ref{lem:
fair pg consistent}, and from the fact that the event structure
constructed in Lemma~\ref{lem:ex fair} satisfies $\ff \FairS(l)$.
\qed
The notion of strong communication fairness is essential for the
results above: $\ff {Fair_I(\varphi,}$ ${\ff t,}$ ${\ff l)}$ may not be realizable if
we assume that communication satisfies only a weak notion of
fairness that says that if a message is sent after some point on,
then it is eventually received. This is so essentially because our
programming language is replacing standard ``if condition then take
action'' programs with weaker variants that ensure that, if after
some point a condition holds, then eventually some action is taken.

We now show that the composition of $\ff{Fair\mbox{-}Pg(\varphi,
t,l,a)}$ and $\ff{Fair\mbox{-}Pg(\varphi, t,l',a)}$ for different
links $l$ and $l'$
satisfies the corresponding fairness assumptions.
\begin{lem}\label{lem:ex fair multiple}
For all distinct
actions $\ff a$ and $\ff a'$,
and
all
distinct links $l$ and $l'$,
${\ff Fair\mbox{-}Pg}(\varphi,$ $  t ,l,$ $a)$ $\oplus $ $ {\ff Fair\mbox{-}Pg}(
\varphi' , t' ,l',a') $
satisfies
$$\begin{array}{l}
\lambda \es. {\ff (\FairS(l)(\es) \land \FairS(l')(\es)) \rimp }\\
\quad {\ff (Fair_I(\varphi,t,l)(\es) \wedge
Fair_I(\varphi',t',l')(\es))}
\end{array}
$$ with respect to
all interpretations $I$ such that $\phi$ is an $i$-formula, $t$ is
an $i$-term, $\phi'$ is an $i'$-formula, and $t'$ is an $i'$-term in
$I$.
\end{lem} \proof Suppose $\ff a \ne \ff a'$. We again use scheme
$\it{GS}$.

{\bf Step 1.} Clearly, we can take $\phi_1$ to be ${\ff \lambda \es.
~\FairS(l)(es)  \rimp Fair_I(\varphi,t,l)(es)}$ and $\phi_2$ to
be ${\ff \lambda \es. ~\FairS(l')(es) \rimp
Fair_I(\varphi',t',l')(es)}$.

{\bf Step 2.} By Lemma~\ref{lem:ex fair}, ${\ff Fair\mbox{-}Pg( }
\varphi {\ff ,}
 {\ff t,l,a)}\psatI
\phi_1$ and ${\ff Fair\mbox{-}Pg(}\varphi'{\ff ,t' ,l',a')} \psatI
\phi_2$. \qed

\noindent Finally, we can
show that ${\ff Fair\mbox{-}Pg(}\varphi{\ff ,t ,l,a)} \oplus  {\ff
Fair\mbox{-}Pg(}\varphi'{\ff ,t' ,l',a')}$ is
{\pgrealizable},
where $l$ is a link from $i$ to $j$, $l'$ is a link from $i'$ to $j'$,
and $l \ne l'$ (so that we may have $i=i'$ or $j=j'$, but not both),
and thus the specification $\lambda \es. {\ff (\FairS(l)(}$
$\ff{\es) \land \FairS(l')(\es))} \rimp $ ${\ff
(Fair_I(\varphi, t,l)(\es) \wedge Fair_I(\varphi', }$ $\ff {t',
l')(\es))}$ is realizable with respect to $I$.
if both $\phi$ and $\phi'$ satisfy the principle of excluded middle
with respect to $I$.

\begin{lem}\label{lem: pg fair plus}
For all interpretations $I$ such that $\phi$ is an $i$-formula, $t$
is an $i$-term, $\phi'$ is an $i'$-formula, and $t'$ is an $i$-term
in $I$,
if both $\phi$ and $\phi'$ satisfy the principle of excluded middle
with respect to $I$, then, for all
distinct actions $a$ and $a'$ and all distinct links $l$ and $l'$, ${\ff
Fair\mbox{-}Pg(}\varphi{\ff ,t ,l,a)} \oplus  {\ff
Fair\mbox{-}Pg(}\varphi'{\ff ,t' ,l',a')}$ is \pgrealizable\ with
respect to $I$. \end{lem}
\proof
If $I(\neg \phi \land \neg \phi')(s)$ holds for some global
state $s$, then let $\es$ be the empty event structure such that
$\initstate_i = s_i$ and $\initstate_{i'} = s_{i'}$.
Clearly $\es$ is consistent with
${\ff
Fair\mbox{-}Pg(}\varphi{\ff ,t ,l,a)} \oplus  {\ff
Fair\mbox{-}Pg(}\varphi'{\ff ,t' ,l',a')}$.
Otherwise, let
$\es$ be an event structure with domain $\Val_I$, $i,j,i',j' \in
\AG$, and $l, l' \in \Links$, consisting of an infinite sequence of states
such that if $I(\phi)$ holds for infinitely many states, then $i$ sends $t$
on link $l$ infinitely often; if $I(\phi')$ holds for infinitely
many states, then $i'$ sends $t'$ on link $l'$ infinitely often; if $t$
is sent on $l$ infinitely often, then $j$ receives it on link $l$
infinitely often; and if $t'$ is sent on $l'$ infinitely often, then
$j'$ receives it on $l'$ infinitely often.  It is straightforward to
construct such an event structure $\es$.  Again, it should be clear that
$\es$ is consistent with
${\ff
Fair\mbox{-}Pg(}\varphi{\ff ,t ,l,a)} \oplus  {\ff
Fair\mbox{-}Pg(}\varphi'{\ff ,t' ,l',a')}$.
\qed

\section{Adding knowledge to Nuprl}\label{sec: k}
We now show how knowledge-based programs can be introduced into
Nuprl.

\subsection{Consistent cut semantics for knowledge}\label{sec: cck}
We want to extend basic programs to allow for tests that involve
knowledge.
For simplicity, we take
$\ff{AG} = \{1,2, \dots, n\}$.
As before, we start with finite sets ${\it P}$ of predicate symbols and
${\it F}$ of function symbols, and close off under conjunction,
negation, and
quantification over non-local variables; but now, in addition,
we also close off under application of the temporal operators ${\kf
\Box}$ and $\kf \diamondsuit$,
and the epistemic operators ${\kf K_i}$, ${\kf i = 1, \ldots,
n}$, one for each process ${\kf i}$.
\commentout{ The ``eventually'' operator $\kf \diamondsuit$ is
defined as the dual of $\kf \Box$ in the standard way: $\kf
\diamondsuit \varphi\tbar \neg \Box \neg \varphi$.}
We again want to define a consistency relation in Nuprl for each
program.  To do that, we first need to review the semantics of knowledge.
Typically, semantics for knowledge is given with respect to a pair
${\kf (r,m)}$ consisting of a run ${\kf r}$ and a time ${\kf m}$,
assumed to be the time on some external global clock (that none of
the processes necessarily
has access to \cite{FHMV}).  In event structures, there is no
external notion of time.  Fortunately, Panangaden and Taylor
\cite{PT} give a variant of the standard definition with respect
to what they call {\em asynchronous runs}, which are essentially
identical to event
structures.
We can simply apply their definition in our framework, replacing using ``event
structure'' instead of ``asynchronous run'', as we do in the following
account.

The truth of formulas is defined relative to a
pair $({\kf Sys, c})$, consisting of a system ${\kf Sys}$ (i.e., a
set of event structures) and a
{\em consistent cut}
${\kf c}$ of some event structure ${\kf es} \in \Sys$, where  a {\em
consistent cut} ${\kf c}$ in
{\kf es} is a set of events in ${\kf es}$ closed under the causality
relation. Recall from Section~\ref{sec:eventsystems} that this
amounts to $\kf c$ satisfying the constraint that, if ${\kf e'}$ is
an event in ${\kf c}$ and ${\kf e}$ is an event in ${\kf es}$ that
precedes ${\kf e'}$ (i.e., ${\kf e} \prec {\kf e'}$), then  ${\kf
e}$ is also in ${\kf c}$.
We write $c \in \Sys$ if $c$ is a consistent cut in some event structure
in $\Sys$.

\begin{figure}[htb]
\includegraphics[width=32pc,height=14pc,
type=eps,ext=.eps,read=.eps]{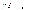} \epsfxsize=17cm
\caption{Two consistent cuts that cannot be
distinguished by agent ${\kf i}$. \label{fig:cck}}
\end{figure}

Traditionally, a knowledge formula $\kf K_i \varphi$ is interpreted
as true at a point $\kf (r,m)$ if $\varphi$ is true regardless of
$\kf i$'s uncertainty about the whole system at $\kf (r,m)$. Since
we interpret formulas relative to a pair $({\kf Sys, c})$, we
need to make precise $\kf i$'s uncertainty at such a pair.
For the purposes of this paper, we assume that each agent keeps
track of all the events that have occurred and involved him (which
corresponds to the assumption that agents have {\em perfect
recall}); we formalize this assumption below. Even in this setting,
agents can be uncertain about what events have occurred in the
system, and about their relative order. Consider, for example, the
scenario in the left panel of Figure~\ref{fig:cck}: agent ${\kf i}$
receives a message from agent ${\kf j}$ (event $\ff e_2$),
then sends a message  to agent ${\kf k}$ ($\ff e_3$), then receives
a second message from agent ${\kf j}$ ($\ff e_6$),
and then performs an internal action ($\ff e_7$). Agent $\ff i$
knows that $\ff send(e_2)$ occurred prior to $\ff e_2$ and that $\ff
send(e_6)$ occurred prior to $\ff e_6$. However, ${\kf i}$ considers
possible that after receiving his message, agent ${\kf k}$ sent a
message to ${\kf j}$ which was received by $\ff j$ before $\ff e_7$
(see the right panel of Figure~\ref{fig:cck}).

In general, as argued by Panangaden and Taylor, agent ${\kf i}$
considers possible any consistent cut in which he has recorded the
same sequence of events. To formalize this intuition, we define
equivalence relations ${\kf \sim_i}$, ${\kf i = 1, \ldots, n}$, on
consistent cuts  by taking ${\kf c \sim_i c'}$ if ${\kf i}$'s
history is the same in ${\kf c}$ and ${\kf c'}$. Given two
consistent cuts ${\kf c}$ and ${\kf c'}$, we say that ${\kf c
\preceq c'}$ if, for each process ${\kf i}$, process ${\kf i}$'s
history in ${\kf c}$ is a prefix of process ${\kf i}$'s history in
${\kf c'}$.
Relative to $({\kf Sys, c})$, agent $\kf i$ considers possible
any
consistent cut $\kf c' \in \Sys$ such that ${\kf
c'\sim_i c}$.%

Since the semantics of knowledge
given here
implicitly assumes that agents have
perfect recall, we restrict to event structures that also satisfy
this assumption.  So, for the remainder of this paper, we restrict
to systems where \emph{
local
states encode histories}, that is, we
restrict to systems $\Sys$ such that, for all event structures $\es,
\es' \in \Sys$, if $e$ is an event in $\es$, $e'$ is an event in
$\es'$,
$\ff \agent(e)=\agent(e')=i$, and $\ff \state~ \before~
e=\state~\before ~e'$,
then $i$ has the same history in both $\es$ and $\es'$.
For simplicity, we guarantee this by assuming
that each agent $i$ has a local variable
$\history_i \in X_i$ that encodes its history.  Thus, we take
$\initstate_i(\history_i) = \bot$ and for all events $e$ associated with
agent $i$, we have $(s~\after~e)(\history_i) = (s~\before
~e)(history_i) \cdot  e$.
It immediately follows that in two global states where $i$ has the same
local state, $i$ must have the same history.
Let $\ff System$ be the set of all such systems.

Recall that events associated with the same agent are totally
ordered. This means that we can associate with every consistent cut
$\kf c$ a global state $s^c$: for each agent $i$, $s^c_i$ is
$i$'s
local
state after the last event $e_i$ associated with $i$
 in $c$ occurs.
Since
local
states encode histories, it follows that
if $s^c_i=s^{c'}_i$, then $c\sim_i c'$. It is not difficult to
see that the converse is also true; that is, if $c\sim_i c'$, then
$s^c_i=s^{c'}_i$.
We also write $s^c \prec s^{c'}$ if $c \prec c'$. In the following,
we assume that all global states
in a system $\Sys$
have the form $s^c$ for some
consistent cut $c$.

Nuprl is sufficiently expressive that epistemic and modal operators can be
defined within it.  Thus, to interpret  formulas with
epistemic operators and temporal operators, we just translate them to
formulas that do not mention them.
Since the truth of an epistemic formula depends not only
on a global state, but on a pair
$(\Sys,c)$, where the consistent cut $c$ can be identified with a
global state in some event structure in $\Sys$, the translated formulas
will need to include variables that, intuitively, range over systems and
global states.  To make this precise, we expand the
language so that it includes
rigid binary predicates $\bCC$ and $\succeq$, a rigid binary function
$\ls$, and rigid constants $\bs$ and $\bSys$.
Intuitively,
$\bs$ represents a global state, $\bSys$ represents a system,
$\bCC(x,y)$ holds if $y$ is a consistent cut (i.e., global state) in
system $x$, $\ls(x,i)$ is $i$'s local state in global state $x$,
and $\succeq$ represents the ordering on consistent cuts defined above.

For every formula that does not mention modal operators, we take $\phi^t
= \phi$.
We define
$$(K_i \phi)^t \eqdef
\forall \bsp ((\bCC(\bSys, \bsp) \land \ls(\bsp,i) = \ls(\bs,i))
\rimp \phi^t[ \bs / \bsp])),$$
$$(\Box \phi)^t \eqdef
\forall \bsp ((\bCC(\bSys,\bsp) \land \bsp \succeq \bs \rimp
\varphi^t[ \bs / \bsp]),$$
and $$(\diamondsuit \phi)^t \eqdef
\exists \bsp ((\bCC(\bSys,\bsp)
\land \bsp \succeq \bs \land \varphi^t[ \bs / \bsp]).$$

Given an interpretation $I$, let $I'$ be the interpretation that extends
$I$ by adding
(i.e., conjoining)
to $\phi_I$
formulas characterizing $\bSys$, $\bs$,
$\bCC$, $\ls$, and $\succeq$ appropriately.  That is, the formulas force
$\bSys$ to represent a set of event structures, $\bs$ to be a consistent
cut in one of these event structures, and so on.
These formulas are all expressible in Nuprl.
We now define a predicate $I'_V(\phi)$ on systems and global states by
simply taking
$I'_V(\phi)(\Sys,s)$ to hold iff
$\phi_{I'}$ together with the
conjunction of atomic formulas of the form $x=V(x)$ for all
non-local variables $x$ that appear in $\phi$,
$x=s_i(x)$ for variables $x \in X_i$, $i \in \AG$, that appear in
$\phi$, $\bs = s$, and  $\bSys = \Sys$, imply $(\phi^t)^+$
(where, in going from $\phi^t$ to $(\phi^t)^+$, we continue to use the
$\bs$).  Thus, we basically reduce a modal formula to a non-modal
formula, and evaluate it in system $\Sys$ using $I_V$.

Just as in the case of non-epistemic formulas,
the valuation $\kf V$ is not needed to interpret formulas
whose only free variables are in $\union_{i\in AG}X_i$.
For such formulas, we typically write $\kf I'(\varphi)(\Sys,s)$ instead of
$I'_V(\varphi)(\Sys,s)$.
We can also define $i$-formulas and $i$-terms in an interpretation $I'$.
For an $i$-formula, we often write $I'_V(\phi)(\Sys,s_i)$ rather than
$I'_V(\phi)(\Sys,s)$.
Note that a Boolean combination of epistemic
formulas whose outermost knowledge operators are $K_i$
is guaranteed to be an $i$-formula in every interpretation, as is a
formula that has no nonrigid functions or predicates and does not
mention $K_j$ for $j \ne i$.  The former claim is immediate from the
following lemma.

\pro\label{pro:ki-formulas} For all formulas $\varphi$, systems
$\Sys$, and
global
states $\ff s$ and $\ff s'$, if $\ff s_i=s'_i$, then
$I'(K_i\varphi)(\Sys,s)$ holds iff $I'(K_i\varphi)(\Sys,s')$ does. \epro

\proof
Follows from the observation that if we have a proof in Nuprl
that an $i$-formula holds given $I'$, $\Sys$, and  $s \in \Sys$,
then we can rewrite the proof so that it  mentions only
$s_i$ rather than $s$.  Thus, we actually have a proof  that
the $i$-formula holds in all states $s' \in \Sys$ such that $s'_i=s_i$.
\qed

\subsection{Knowledge-based programs and specifications}\label{sec:kb}
In this section, we show how we can extend the notions of program
and specification presented in  Section~\ref{sec:overview} to
knowledge-based programs and specifications. This allows us to
employ the large body of tactics and libraries already developed in
Nuprl to synthesize knowledge-based programs from knowledge-based
specifications.

\subsubsection{Syntax and semantics}\label{sec:kb syntax}

Define \emph{knowledge-based message automata} just as we defined
message automata in Section~\ref{sec:dma}, except that we
now allow arbitrary epistemic formulas in tests.  If we want to
emphasize that the tests can involve knowledge, we talk about
\emph{knowledge-based} initialization, precondition, effect, and
fairness programs.
For the purposes of this
paper, we take knowledge-based programs to be knowledge-based
message automata.
Formally, there are five basic knowledge-based clauses for
agent $i$:
\begin{enumerate}[$\bullet$]
\item ${\cf @} i~{\cf initially~ \psi}$;
\item ${\cf @} i~ {\cf if}~\kind =  k~{\cf then}~ x:= t$;
\item  ${\cf @} i~ \kind= \local(a)~{\cf only~if~} {\cf\phi}$;
\item ${\cf @} i~{\cf if~necessarily}~\phi~{\cf
then~i.o.}~\kind=\local(a) $; and
\item ${ \cf @} i~ {\cf only~events~in}~ L~ {\cf affect}~ x$,
\end{enumerate}
where $\psi$ and $\phi$ are $i$-knowledge-based formulas, $k \in \Act
\union {\ff Links}$,
$x \in X_i$, $t$ is an $i$-term, and $\ff L$ is a list of  kinds in $\Act \union Links$.

We give semantics to knowledge-based programs by first associating with
each know\-ledge-based program a function from systems to systems.
Let $(\Pg^{kb})^t$ be the result of replacing every formula $\phi$ in
$\Pg^{kb}$ by $\phi^t$.
Note that $(\Pg^{kb})^t$ is a standard program, with no modal formulas.
Given an interpretation $I$ and a system
$\Sys$, let $I(\Sys)$ be the interpretation that is characterized by
the formula that results from adding
(i.e., conjoining)
to $\phi_{I}$ the
formula $\bSys = \Sys$.%
\footnote{The notation $I(\Sys)$
may seem somewhat awkward for a formula, but in this case it is a
formula that characterizes a system, so it is perhaps not so
unreasonable.  In any case, since this formula will appear in subscripts
(e.g., in Definition~\ref{def:kb sem}), it seems a better choice than,
say, $I_\Sys$.}
Now we can apply the semantics of Section~\ref{sec:sem} to get the
system
$S_{I(\Sys)}((\Pg^{kb})^t)$.
In more detail, since $(\Pg^{kb})^t$ is a standard program, we can apply
Definition~\ref{def:sem},
which says that the semantics of $(\Pg^{kb})^t$
with respect to the interpretation $I(\Sys)$
is the set of all event structures in $\Sys$ that are
consistent
with $(\Pg^{kb})^t$ with respect to $I(\Sys)$; that is,
$$S_{I(\Sys)} ((\Pg^{kb})^t) = \{ \mathit{es}~ | ~ \mathit{Consistent}_{I(\Sys)}((\Pg^{kb})^t, \mathit{es} ) \}.$$
(Note that, technically, $(Pg^{kb})^t$ does take a system as an argument, which was not the case
of the type of programs defined in Section 2.; however, as $(Pg^{kb})^t$ is independent of the system argument,
its semantics is also independent of any system, which is why we treat it as a standard program.)
Since $(\bSys = \Sys)$ is a conjunct of $I(\Sys)$,
all the event structures $\mathit{es}$ in $S_{I(\Sys)}((\Pg^{kb})^t)$ must be in the set $\Sys$; in other words,
$S_{I(\Sys)} ((\Pg^{kb})^t) \subseteq \Sys$, for all systems $\Sys$.

In general, the system
$S_{I(\Sys)}((\Pg^{kb})^t)$ will be a strict subset of the system
$\Sys$.
Indeed, $S_{I(\Sys)}((\Pg^{kb})^t)$ may even be empty (if there exists no
event structure in $\Sys$ consistent with $(Pg^{kb})^t$ when interpreted
with respect to $I(\Sys)$).
For example, consider a system with two agents, $i$ and $j$, where
$x_j \in X_j$ is a local
variable of agent $j$. Let $i$ follow the simple program
$\Pg^{kb} = @i ~{\bf{initially}}~ K_i(x_j=0)$
that says that initially $i$ knows that $j$'s variable $x_j$ has value
$0$.  Clearly, $\Pg^{kb}$ is a knowledge-based program, an instance of
the knowledge-based initialization
clause $@i ~{\bf{initially}}~\psi$ for $\psi = K_i(x_j=0)$.
By definition, $\psi^t = \forall {\bf{s'}}.~
( {\bf CC} (\bSys, {\bf{s'}}) \wedge
{\bf{ls}}({\bf{s'}}, i)={\bf{ls}}({\bf{s}}, i)) \rimp({\bf{s'}}_j(x)=0)$. That is,
$I(\Sys)(\psi^t)(\initstate_i) = (\bSys = \Sys) \wedge
\forall {\bf{s'}}.~ ( {\bf CC} (\bSys, {\bf{s'}}) \wedge
{\bf{ls}}({\bf{s'}}, i)={\bf{ls}}({\bf{s}}, i)=\initstate_i \rimp({\bf{s'}}_j(x)=0)$.
This means that an event structure $es$ in $\Sys$ is consistent with the clause
$@i ~{\bf{initially}}~ \psi^t$ (i.e., with $(\Pg^{kb})^t$) if and only
if, for all consistent cuts (i.e., global states) $s$ in $\Sys$ such that
$i$'s local state in $s$
is same as the initial state of $i$ in $es$, $x_j$ has value $0$ in $j$'s local state
in $s$. Consider now a system $\Sys$ such that $s_j(x_j)\neq 0$ for all $s$ in $\Sys$.
Clearly, no event structure in $\Sys$ satisfies this condition,
which means that
$S_{I(\Sys)}((\Pg^{kb})^t) = \emptyset$. On the other hand,
if $Sys$ is a
system such that $s_j(x_j) = 0$ for all $s$ in $\Sys$, then
$S_{I(\Sys)}((\Pg^{kb})^t) = \Sys$.

A system $\Sys$ {\em represents} a knowledge-based program
$\ff{ Pg^{kb}}$
(with respect to interpretation $I$) if it is a fixed point of this
mapping; that is,
if $S_{I(\Sys)}((\Pgkb)^t) = \Sys$.
Intuitively, if $\Sys$ is a fixed point, then when interpreted with
respect to $\Sys$, the program is acting the way it should.
Following Fagin et al.~\cite{FHMV,FHMV94}, we take the
semantics of a knowledge-based program $\Pgkb$ to be the set of
systems that represent it.

\dfn\label{def:kb sem}
A \emph{knowledge-based program semantics} is a function associating
with a knowledge-based program $\ff{Pg^{kb}}$
and an interpretation $I$
the systems that represent $\ff{Pg^{kb}}$
with respect to $I$; that is, $\ff S^{kb}_{I}(\Pgkb)=\{ Sys \in
\mathit{System} \:
| \:
S_{I(\Sys)}((\Pgkb)^t) = \Sys\}$. \edfn

As observed by Fagin et al.~\cite{FHMV,FHMV94}, it is
possible to construct knowledge-based programs that are represented
by no systems, exactly one system, or
more than one system.  %
However, there exist
conditions (which are often satisfied in practice) that guarantee
that a knowledge-based program is represented by exactly one system.
Note that, in particular, standard programs, when viewed as
knowledge-based programs,
are represented by a unique system; indeed, $\ff{ S^{kb}_{I}(Pg)} =
\{\ff{ S_I(Pg)}\}$.  Thus,  we can view $\ff{ S^{kb}_{I}}$ as
extending $\ff{ S_I}$.
A (standard) program $\ff Pg$ {\em implements} the knowledge-based
program $\Pgkb$ with respect to interpretation $I$ if $S_I(\Pg)$
represents $\Pgkb$ with respect to $I$, that is, if
$S_{I(S_I(\Pg))}((\Pgkb)^t)$ $=$ $S_I(\Pg)$.
In other words, by interpreting the tests in $\Pg^{kb}$ with respect
to the system generated by $\Pg$, we get back the program $\Pg$.

\subsubsection{Knowledge-based specifications}\label{sec:kb specs}
Recall that a standard specification is a predicate on event
structures.
Following \cite{FHMV94}, we take a knowledge-based specification
to be a predicate on systems.
\dfn\label{def:satk} A
\emph{knowledge-based specification} is a predicate on
$\mathit{System}$.
A knowledge-based program $\ff{ Pg^{kb}}$ satisfies a
knowledge-based specification $\ff{ Y^{kb}}$ with respect to $I$,
written $\ff{ Pg^{kb} \psatkbI Y^{kb}}$, if all the systems
representing $\ff{ Pg^{kb}}$ with respect to $I$ satisfy $\ff{
Y^{kb}}$, that is, if the following formula holds: $\ff{ \forall
Sys\in S^{kb}_{I} (Pg^{kb}).~ Y^{kb}(Sys)}$. The knowledge-based
specification $\ff Y^{kb}$ is {\em realizable} with respect to $I$
if there exists a (standard) program $\ff Pg$ such that $\ff
S_I(Pg)\neq \emptyset$  and $\ff Pg\psatkbI Y^{kb}$ (i.e., $\ff
Y^{kb}(S_I(Pg))$ is true). \edfn

\commentout{ As for standard basic clauses, it is not difficult to
show that all knowledge-based clauses, except initialization ones,
are trivially {\pgrealizable}: we simply take $\Sys$ to consist of
only one event structure $\es$ with no events. In the case of
initialization clauses ${\cf @}{\ff i}~{\cf initially}~ \psi$, if
$\psi$ is not invalid and $\mathit{Determinate}(\psi)$ holds, we can
take $\Sys$ to consist of one event structure $\es$ with no events
and $\Sigma_i= \lbrace s_i \rbrace$ such that $\I'(\psi)(\Sys, s_i)$
holds.}
As for standard basic programs, it is not difficult to show that
knowledge-based precondition, effect, and frame programs are
trivially {\pgrealizable}: we simply take $\Sys$ to consist of only
one event structure $\es$ with no events.
A
knowledge-based initialization
program is realizable iff $\phi_{I} \land \psi^t$ is satisfiable.
Finding sufficient conditions for fair knowledge-based
programs to be realizable is nontrivial. We cannot directly
translate the constructions sketched for the standard case to the
knowledge-based case because, at each step in the construction (when
an event structure has been only partially constructed), we would
have to argue that a certain knowledge-based fact holds when
interpreted with respect to an entire system and an entire event
structure. However, in the next section, the knowledge-based
programs used in the argument for STP (which do include fairness
requirements) are shown to be realizable.

\subsubsection{Axioms}\label{sec:kb axioms}
We now consider the extent to which we can generalize the axioms
characterizing (standard) programs %
presented in Section~\ref{sec:dma} to knowledge-based programs.

Basic knowledge-based message automata other than knowledge-based
precondition and fairness requirement programs satisfy analogous
axioms to their standard counterparts.  The only difference is that
now we view the specifications as functions on systems, not on event
structures. For example, the axiom corresponding to $\bf
Ax\mbox{-}init$ is
$$
\begin{array}{l}
{\bf Ax\mbox{-}initK}:\\
\quad  {\cf @}{\ff i}~{\cf initially}~ \psi \ff{ \,
\psatkbI }
\ff{\lambda \Sys.~i\mbox{-\emph{formula}}(\psi,I)\land \forall
es\in
                       \Sys.~I(\psi)(\Sys, \initstate_i)}.
\end{array}
$$
(Note that here, just as in the definition of $\bf Ax\mbox{-}init$,
for simplicity, we write $\initstate_i$ instead of
$\es.\initstate_i$. Since $\psi$ is constrained to be an $i$-formula in
$I$, it
makes sense to talk about $I(\psi)(\Sys, \initstate_i)$ instead of
$I(\psi)(\Sys, s)$ for a global state $s$ with $s_i =
\initstate_i$.)
The knowledge-based  analogues of  axioms $\bf Ax\mbox{-}cause$,
 $\bf Ax\mbox{-}affect$, and $\bf Ax\mbox{-}sends$ are denoted $\bf
Ax\mbox{-}causeK$, $\bf Ax\mbox{-}affectK$, and $\bf
Ax\mbox{-}sendsK$, respectively,
and are identical to the standard versions of these axioms.
The knowledge-based counterparts of
${\bf Ax\mbox{-}if}$  and ${\bf Ax\mbox{-}fair}$ now involve
epistemic preconditions, which are interpreted  with respect to a
system:
$$
\begin{array}{ll}
{\bf Ax\mbox{-}ifK}: & {\cf @}i~\kind=\local(a) ~{\cf
only~if}~\varphi \psatkbI
~ \lambda \Sys.~ i\mbox{-\emph{formula}}(\phi,I) \land \\
& \forall \es\in \Sys.~\forall e@i\in
\es.~(\kind(e)=\local(a))\Rightarrow I(\varphi)(\Sys,\state~\before~e)\\
\end{array}
$$
$$
\begin{array}{ll}
{\bf Ax\mbox{-}fairK}:&
  {\cf @}i~{\cf if~necessarily}~\varphi~{\cf then ~i.o.}~\kind=\local(a)
  \, \psatkbI {\ff \lambda \Sys. ~i\mbox{-\emph{formula}}(\phi,I) \land}\\
  & \begin{array}{ll}
\forall \es \in \Sys. &
  ( (\exists e@i \in \es \land \forall e@i\in
      \es.~ \exists e'\essucceqi
    e.~ \\
    & \quad I(\neg \phi)(\Sys,\state~\after~e') \vee \kind(e')=local(a)) \vee \\
     & ~(\neg(\exists e@i \in \es) \land  I(\neg \phi)(\Sys,
    {\initstate}_i(\es)))).
    \end{array}\\
\end{array}
$$

There are also obvious analogues axioms ${\bf Ax\mbox{-}ref}$ and
${\bf Ax\mbox{-}\oplus }$, which we denote ${\bf Ax\mbox{-}refK}$
and ${\bf Ax\mbox{-}\oplus K}$ respectively. \begin{lem}\label{lem:kb pg}
Axioms $\bf{Ax\mbox{-}initK}$, $\bf{Ax\mbox{-}causeK}$,
$\bf{Ax\mbox{-}affectK}$, $\bf{Ax\mbox{-}sendsK}$, ${\bf
Ax\mbox{-}ifK}$,
${\bf Ax}$ ${\bf \mbox{-}}$ ${\bf fairK}$,
and ${\bf Ax\mbox{-}refK}$ hold for all interpretations $I$. \end{lem}
\proof Since the proofs for all axioms are similar in spirit, we prove
only
that $\bf Ax\mbox{-}ifK$ holds for all interpretations
 $I'$.  Fix an interpretation $I$. Let $\ff Pg^{kb}$ be the program ${\cf @}i
~\kind=\local(a)~{\cf only~if}~\varphi$,
where $\phi$ is an $i$-formula.
Let $\ff Y^{kb}$ be
an instance of $\bf Ax\mbox{-}ifK$:
\commentout{
$$
\begin{array}{l}
{\ff \lambda \Sys.~\forall \es\in \Sys.}  \\
 \quad
\ff{\forall e@i\in \es.~(\kind(e)=\local(a) \Rightarrow
 P_{\Sys,\varphi}(\val(e),\state~\before~e))}\\
 \quad  \ff{\wedge ~\forall e@i\in \es.~ \exists e'>_i e.~
 (\kind(e')=\local(a)~\vee ~
\forall v'.~ \neg P_{\Sys,\varphi}(v',\state~\after~e'})).\\
\end{array}
$$
}
$$
\begin{array}{l}
\lambda \Sys.~ i\mbox{-\emph{formula}}(\phi,I) \land \forall \es\in \Sys. ~\forall e@i\in \es.~
                             (\kind(e)=\local(a))
                  \Rightarrow \\
\quad I(\varphi)(\Sys,\state~\before~e).
\end{array}
$$
By Definition~\ref{def:satk}, ${\ff Pg^{kb}}\psatkbI \ff{Y^{kb}}$ is
true if and only if, for all systems
$\ff Sys \in S^{kb}_{I}(Pg^{kb})$,
$\ff Y^{kb}(Sys)$ holds.
That is, for all systems $\ff \Sys$ such that $\ff
S_{I(\Sys)}((\Pgkb)^t)=\Sys$, the following holds:
\commentout{
$$
\begin{array}{ll}
{\ff \forall \es\in \Sys.}
  &
 {\ff \forall e@i\in \es.~(\kind(e)=\local(a) \Rightarrow
  P_{\Sys,\varphi}(\val(e),\state~\before~e))}\\
  &
\ff{\wedge ~\forall e@i\in \es.~ \exists e'>_i e.~
(\kind(e')=\local(a)~\vee~
 \forall v'.~ \neg P_{\Sys,\varphi}(v',\state~\after~e'})).\\
\end{array}
$$
}
$$\forall \es\in \Sys. ~i\mbox{-\emph{formula}}(\phi,I)\land \forall e@i\in \es.~
                             (\kind(e)=\local(a))
                  \Rightarrow I(\varphi)(\Sys,\state~\before~e).$$
Let $\ff \Sys$ be a system such that $\ff
S_{I(\Sys)}((\Pgkb)^t)=\Sys$.
By Definition~\ref{def:sem},
all event structures in $\ff \Sys$
are consistent with the program $\ff (\Pgkb)^t$ with respect to
interpretation $I(\Sys)$. Recall that $\ff (\Pgkb)^t$ is the
(standard) program
${\cf @}i ~\kind=\local(a)~{\cf only~if}~{\cf {\phi}^t}$,
where $I(\Sys)({\phi}^t)(s)=I(\phi)(\Sys,s).$
We can thus apply axiom ${\bf Ax\mbox{-}if}$ and conclude that the
following holds for all event structures
$\ff \es$ consistent with $\ff I(\Sys)((\Pgkb)^t)$ with respect to
$I(\Sys)$
(i.e., for all $\ff \es\in \Sys$):
\commentout{
$$
\begin{array}{l}
{\ff \forall e@i\in \es.~(\kind(e)=\local(a) \Rightarrow
P_{\Sys,\varphi}(\val(e),\state~\before~e))}\\
\ff{\wedge ~\forall e@i\in \es.~ \exists e'>_i e.~
(\kind(e')=\local(a)~\vee~  \forall v'.~ \neg
P_{\Sys,\varphi}(v',\state~\after~e'})).\\
\end{array}
$$
}
$$i\mbox{-\emph{formula}}({\phi}^t, I(\Sys)) \land \forall e@i\in \es.~(\kind(e)=\local(a))
             \Rightarrow I(\Sys)({\phi}^t)(\state~\before~e).$$
The first conjunct says that, for all global states $s$ and $s'$ in
$\Sys$, if $s_i=s'_i$ then
$I(\Sys)({\phi}^t)(s)$ $=$ $ I(\Sys)$ $({\phi}^t)(s')$, which is equivalent
to saying that $I(\phi)(\Sys,s)=I(\phi)(\Sys,s')$, that is,
$i\mbox{-\emph{formula}}(\phi,I)$ holds. The second conjunct is
equivalent to
$$\forall e@i \es. ~ (\kind(e)=\local(a)) \rimp I(\phi)(\Sys, \state~\before~e),$$
by the definition of ${\phi}^t$ and $I(\Sys)$.
Thus, $\ff Y^{kb}(\Sys)$ holds
under interpretation $I$.
\qed

The  proof of Lemma~\ref{lem:kb pg} involves only unwinding the
definition of
satisfiability for knowledge-based specifications and
the application of simple refinement rules, already implemented in
Nuprl. In general, proofs of epistemic formulas will also involve
reasoning in the logic of knowledge. Sound and complete
axiomatizations of (nonintuitionistic) first-order logic of
knowledge are well-known (see \cite{FHMV} for an overview) and
can be formalized in Nuprl in a straightforward way. This is
encouraging, since it supports the hope that Nuprl's  inference
mechanism is powerful enough to deal with knowledge specifications,
without further essential additions.

\commentout{ The following simple result reduces the problem of
finding a program that satisfies a knowledge-based specification to
finding a knowledge-based program that satisfies the specification,
and then constructing a (standard) program that implements the
knowledge-based program: \rem\label{rem: impl} If $\ff{Pg^{kb}\psat
Y^{kb}}$ and $\ff{Pg}$ implements $\ff{Pg^{kb}}$, then $\ff{Pg\psat
Y^{kb}}$. \erem
We succinctly write this as the axiom
$$
\begin{array}{ll}
\bf{Ax\mbox{-}imp}:& \ff{(Pg^{kb}\psat Y^{kb})~\wedge
~(Pg~implements~Pg^{kb}) \Rightarrow (Pg\psat Y^{kb})}
\end{array}
$$
By the definition of satisfiability, the following refinement axiom
can be proved:
$$
\begin{array}{ll}
\bf{Ax\mbox{-}RefK}: & \ff{(Pg^{kb}\psat Y^{kb})~\wedge~(
Y^{kb}\Rightarrow Y^{kb})\Rightarrow (Pg^{kb}\psat Y^{kb})}
\end{array}
$$
}
Note that {\bf Ax}\mbox{-}$\oplus${\bf K} is not included in
Lemma~\ref{lem:kb pg}.  That is because it does not always hold, as
the following example shows. \xam\label{ex:oplusK}
Let $\ff Y_i^{kb}\eqdef \Box (\neg K_{1-i}(x_i=i))$ for $i = 0,1$,
where $x_i \in X_i$,
and let $I = \emptyset$.
Let $\ff Pg_i$, $i = 0,1$ be the standard
program for agent $i$ such
that $S_I(\Pg_i)$ consists of all the event structures such that
$x_i = i$ at all times; that is, $\Pg_i$ is the program
$${\cf @} i~{\cf initially}~\ff{x_i=i}~\oplus ~
             {\cf @}i~{\cf only~events~in}~\ff{\emptyset}~\cf{affect}~\ff{x_i}.$$
Since $\Pg_i$ places no constraints on $\ff x_{1-i}$,
is straightforward to prove that $\ff Pg_i\psatkbI Y_{1-i}^{kb}$,
for $i = 0, 1$.  On the other hand,
$S_I(\Pg_1\oplus \Pg_2)$ consists of all the event structures where
$x_i = i$ at all times, for $i = 0, 1$,
so $\ff Pg_0\oplus Pg_1 \psatkbI \neg Y_0^{kb} \land \neg Y_1^{kb}$.
\exam

\subsection{Examples}
In this section, we give some examples of programs in our framework.  A
few simple programs are given in Section~\ref{sec:simplexam},
while a more complex program is discussed in Section~\ref{example: kb
spec}.

\subsubsection{Simple examples}\label{sec:simplexam}
Suppose that a sender $S$ wants to send the value of a bit to a receiver
$R$, and that this value does not change over time.  This
can be easily modeled in our
framework by requiring the sender $S$ to follow this program:
$$
\begin{array}{l}
\ff{@S~{\cf initially}~(x_S =0)\vee (x_S=1)} ~\oplus \\
\ff{@S~\mbox{\cf{only~events~in~}} [ ~] ~\mbox{\cf{affect}}~x_S,}
\end{array}
$$
where $x_S$ is a variable local to $S$. The first clause says that
the initial value of $x_S$ is either $0$ or $1$, while the second
clause says that the value of $x_S$ does not change.

Call this program $\Pg_S^0$. The goal is for the
receiver $R$
to eventually know the (value of the) bit $x_S$. We write this
specification simply
as $Y^{kb}\eqdef \diamondsuit K_R(x_S)$, where $K_R(x_S)$ is an
abbreviation for $K_R(x_S =0)\vee K_R(x_S=1)$. Intuitively, whether
this specification is satisfiable or not depends on the assumptions
made regarding the communication between $S$ and $R$, that is, regarding
the links
$l_{SR}$ and $l_{RS}$, and on whether agents forget facts they once
knew. For simplicity, we assume that agents have perfect
recall. Among other things, this implies that if $R$ knows the bit at some
point in time, since the bit does not change its value, $R$ will
know the the value of the bit at all later times. Suppose we further
assume that communication on $l_{SR}$ is reliable: all messages sent
on $l_{SR}$ are guaranteed to be eventually received by $R$. It is
then
not difficult to see that $Y^{kb}$ is achieved if the sender
continues to send
the bit to $R$ as long as he does not know that $R$ knows the bit.
For if at some point in time  $S$ knows
that $R$ knows the bit, then $R$ knows the bit, and will
subsequently always know it; and if $S$ does not know that $R$ knows
the bit, then $S$ will send the value of the bit and $R$ will
eventually receive it. We can model this in the framework by
assuming that $S$ follows the program
$$
\begin{array}{ll}
\ff{Pg_S^0 } & ~\oplus ~\ff{@S~\kind=\local(a_S)~\mbox{\cf{only
~if}}\ \neg K_S (K_R (x_S))} \\
& ~\oplus~ \ff{@S~\mbox{\cf{if}}~\kind=\local(a_S)~\mbox{\cf{then}}~\msg(l_{SR}) ~:=~x_S}.
\end{array}
$$

The role of $R$ so far has been passive. If the communication on $l_{RS}$ is
also reliable,
we can ensure that $S$ sends fewer messages by having $R$ sending some token to $S$ as soon as
he receives the bit. To reason at a more abstract level, we can ensure that $R$ sends a token to $S$
as soon as he knows the bit. This is modeled by having $R$ follow the program:
$$
\begin{array}{l}
\ff{@R~\kind=\local(a_R)~\mbox{\cf{only ~if}}~\ K_R (x_S)} ~\oplus \\
\ff{@S~\mbox{\cf{if}}~\kind=\local(a_R)~\mbox{\cf{then}}~\msg(l_{RS}) ~:=~ \it{token}},
\end{array}
$$
where $\it{token}$ is an arbitrary constant. In this program, $a_R$ is the action of $R$ sending a token to $S$.

With this program, $R$ continues to send the token once he learns the
bit.  We can minimize communication further by having $R$ send the token
only if he does not
know that $S$ knows that he knows the bit:
$$
\begin{array}{l}
\ff{@R~\kind=\local(a_R)~\mbox{\cf{only ~if}}~\ K_R (x_S) \wedge \neg K_R (K_S K_R (x_S))} ~\oplus \\
\ff{@S~\mbox{\cf{if}}~\kind=\local(a_R)~\mbox{\cf{then}}~\msg(l_{RS}) ~:=~ \it{token}}.
\end{array}
$$

\subsubsection{A knowledge-based specification and program for fairness}\label{example:
kb spec}
Recall from Section~\ref{example: spec}
that the specification $\ff{ \FairS(l) \rimp Fair_I(\varphi, t, l)}$
is
satisfied by the program $\ff {Fair}$ $\mbox{-}$ $\ff {Pg(}\varphi {\ff , t,}$
 $\ff{l,a)}$, for all actions $a$.
We now
consider a knowledge-based version of this specification.
If $\phi$ is an $i$-knowledge-based formula
and $t$ is an $i$-term
in $I$,
define $${\ff Fair^{kb}_I(\phi,t,l)\eqdef \lambda \Sys.~\forall
\es\in \Sys.~
Fair_{I(\Sys)}({\phi}^t,t,l)(\es)},$$ that is
\commentout{
$$
\begin{array}{l}
{\ff Fair^{kb}(P^{kb},f,l) \eqdef}\\
 {\ff \lambda \Sys.~\forall \es\in \Sys.~
      \forall e'\in es.~(\kind(e')=\rcv(l))\Rightarrow } \\
\quad \quad {\ff P^{kb}(\Sys, \val(\send(e')),\state
~\before~\send(e'))}
    \wedge  {\val(e')=f(\val(\send(e')),\state~\before~\send(e'))}\\
\quad  \wedge ~ {\ff \forall e@i\in es.} {\ff \exists e'.~
         (\send(e') \essucci  e \land \kind(e')=\rcv(l)) }
 \vee
     ~(e' \essucceqi e \land \forall v'. ~\neg P^{kb}(\Sys, v',\state~\after~e')).
\end{array}
$$
}
$$
\begin{array}{lll}
{\ff Fair^{kb}_I(\phi,t,l)} \eqdef\\
\quad \quad \lambda \Sys. i\mbox{-\emph{formula}}(\phi,I) \land
i\mbox{-\emph{term}}(t,I)\land \\
\quad \quad \quad \ff{\forall \es\in \Sys.}
                     (({\ff \forall e'\in es.~(\kind(e')=\rcv(l))\Rightarrow } \\
               \quad \quad \quad  \quad {\ff I(\phi)(\Sys,\state ~\before~\send(e'))}
                         \wedge
                    {\ff \val(e')= I(t)(\Sys,\state~\before~\send(e'))})\\
 \quad \quad \quad \quad \wedge  ( (\exists e@i \in \es \land \forall e@i \in \es.~\exists e' \essucceqi e.~
 I(\neg \phi)(\Sys, \state~\after ~e')) \vee \\
\quad \quad \quad \quad  \quad  (\exists e@i \in \es \land \forall e@i \in \es.~\exists e'.~
                    \kind(e')=\rcv(l) \land \send(e') \essucceqi e) \vee\\
\quad \quad \quad \quad  \quad (\neg (\exists e@i\in \es) \land I(\neg \phi)(\Sys, {initstate}_i))).
\end{array}
$$
For example, $\ff Fair^{kb}_I(K_i\varphi, t, l)$ says that every
message received on $l$ is given
by the term $\ff t$ interpreted at the  local state of the sender
$i$, and that $i$ must have known fact $\varphi$ when it sent this
message on $l$;
furthermore, if from some point on $i$ knows that
$\varphi$ holds, then eventually a message is received on $l$.

\commentout{ In Section~\ref{sec:kb}, for an $i$-formula $\phi$
whose only free variable is the value variable $v$, we defined the
predicate $P_{\Sys,\varphi}$ on values and states such that
$$\ff{P_{\Sys,\varphi}(v,s)=T_{\gamma}(\Sys,\es,c,\varphi[v/\val])}.$$
Define $P_\phi$ to be the predicate on systems, values, and states
such that $$P_{\varphi}(\Sys,v,s)=P_{\Sys,\varphi}(v,s).$$ }
\commentout{
$$
\begin{array}{l}
{\ff Fair^{kb}(P_{K_i\varphi},f,l) \tbar \lambda Sys.~\forall es\in Sys.}\\
 \begin{array}{ll}

                          {\ff \forall e'\in es.~(\kind(e')=\rcv(l))\Rightarrow } &
                                      {\ff P_{K_i\varphi}(\Sys,\val(\send(e')),\state ~\before~\send(e'))}\\
                         & \wedge \\
                                                                                 &
                                      {\val(e')=f(\val(\send(e')),\state~\before~\send(e'))}\\
                         {\ff \wedge } &\\
 \end{array}\\
 \begin{array}{ll}
                         {\ff \forall e@i\in es.} &
                                          {\ff \exists e'.~(\kind(e')=\rcv(l))\wedge (\send(e')>_i e)}\\
                         & {\ff \vee }\\
                         & {\ff \exists e'>_i e.~\forall v'. ~
                                    \neg P_{K_i\varphi}(\Sys,v',\state~\after~e')}.
 \end{array}
\end{array}
$$
}
As in Section~\ref{example: spec}, we assume that message
communication satisfies a strong fairness  condition. The
knowledge-based version of the condition $\FairS(l)$ simply
associates with each system $\ff Sys$ the specification $\ff
\FairS(l)$; that is, ${\ff \FairS^{kb}(l)}$ is just ${\ff \lambda
\Sys. ~\forall \es\in Sys.} \FairS(l)(\es)$.

\begin{lem}\label{lem:ex fair k}
For all interpretations $I$ such that
$\phi$ is an $i$-formula
and $t$ is an $i$-term
in $I$,
and all actions $\ff a$, we have that
$${\ff Fair\mbox{-}Pg(\varphi}{\ff , t, l,a) \psatkbI
\FairS^{kb}(l)\Rightarrow
Fair^{kb}_{I}(\varphi, t, l)}.$$ \end{lem}

The proof is similar in spirit to that of Lemma~\ref{lem:kb pg}; by
supplying a system $\Sys$ as an argument to the specification, we
essentially reduce to the situation in Lemma~\ref{lem:ex fair}. We
leave details to the reader.
\commentout{ Note that, unlike Lemma~\ref{lem:ex fair}, we do not
have to make an explicit assumption of determinacy here.  This is
because, by the semantics of knowledge formulas,
$\ff \forall s.~\Dec(\exists v.~I'(\varphi)(\Sys,v,s))$  is true for
all  $i$-formulas $\varphi$ and systems $\Sys$.
}%

We can also prove the following analogue of Lemma~\ref{lem:ex fair
multiple}.
\begin{lem}\label{lem:ex fair multiple k}
For all interpretations $I$ such that
$\phi$ is an $i$-formula,
$\phi'$ is a $j$-formula,
$t$ is an $i$-term, and $t'$ is a $j$-term
in $I$,
all distinct links $\ff l$ and $\ff l'$,
and
all
distinct actions $a$ and $a'$, we have that
$$
\begin{array}{l}
{\ff Fair\mbox{-}Pg(\varphi} {\ff , t, l,a)}\oplus
{\ff Fair\mbox{-}Pg(\phi'} {\ff ,t', l',a')} \psatkbI \\
 \quad {\ff (\FairS^{kb}(l) \land \FairS^{kb}(l'))} \rimp
       {\ff (Fair^{kb}_I(\varphi, t, l)\wedge Fair^{kb}_I(\phi', t',l'))}.
\end{array}
$$
\end{lem}

\section{The sequence-transmission problem (STP)}\label{sec:stp}
In this section, we give a more detailed example of how a program
satisfying a knowledge-based specification $X$  can be extracted
from $X$ using the Nuprl system.
We do the extraction in two stages.  In the first stage, we use
Nuprl to prove that the specification is satisfiable. The proof
proceeds by refinement: at each step, a rule or tactic (i.e., a
sequence of rules invoked under a single name) is applied, and new
subgoals are generated; when there are no more subgoals to be
proved, the proof is complete. The proof is automated, in the sense
that subgoals are generated by the system upon tactic invocation.
From the proof, we can extract a knowledge-based program $\ff{
Pg^{kb}}$ that satisfies the specification. In the second stage, we
find standard programs that implement $\ff{ Pg^{kb}}$. This
two-stage process has several advantages:
\begin{enumerate}[$\bullet$]
\item A proof carried out to derive $\ff{ Pg^{kb}}$
does not rely on  particular assumptions about how knowledge is
gained.
Thus, it is potentially more intuitive and elegant than a proof
based on certain implementation assumptions.
\item By definition, if $\ff{ Pg^{kb}}$ satisfies a
specification, then so do all its implementations.
\item This methodology gives us a general technique for
deriving standard programs that implement the knowledge-based
program, by finding stronger (non-knowledge-based) predicates that
imply the knowledge preconditions in $\ff{Pg^{kb}}$.
\end{enumerate}

\noindent We illustrate this methodology by applying it to a problem
that has received considerable attention in the context of
knowledge-based programming, {\em the sequence-transmission problem}
(STP).

\subsection{Synthesizing a  knowledge-based program for STP}\label{sec:kbstp}
The STP involves  a sender $\ff{ S}$ that
 has an input tape with a (possibly infinite) sequence
$\ff{ X= X(0), X(1), \dots }$ of bits, and wants to transmit $\ff{
X}$ to a receiver $\ff{ R}$; $\ff{ R}$ must write this sequence on
an output tape $\ff{ Y}$. (Here we assume that $\ff X(n)$ is a bit
only for simplicity; our analysis of the STP does not essentially
change once we allow $\ff X(n)$ to be an element of an arbitrary
constructive domain.)
A solution to the STP must satisfy two conditions:
\begin{enumerate}[(1)]
\item (safety): at all times, the sequence $\ff{ Y}$ of bits written
by $\ff{ R}$ is a prefix of $\ff{ X}$, and
\item (liveness): every bit $\ff{ X(n)}$ is eventually written by $\ff{
R}$ on the
output tape.
\end{enumerate}
Halpern and Zuck~\cite{HZ} give two knowledge-based programs
that solve the STP, and show that a number of standard programs in
the literature, like Stenning's~\cite{St}
protocol, the alternating-bit protocol~\cite{BSW}, and Aho,
Ullman and
Yannakakis's  algorithms~\cite{AUWY}, are all particular
instances of these programs. \commentout{ Sanders
\cite{sanders_predtrans} derives a number of knowledge-based and
standard programs for the same problem, with a  focus on the more
practical aspects of program development. Our method uses ideas from
both of these earlier works. }

If messages cannot be lost, duplicated, reordered, or corrupted,
then $\ff{ S}$ could simply send the bits in $\ff{ X}$ to $\ff{ R}$
in order. However, we are interested in solutions to the STP in
contexts where communication is not reliable.
It is easy to see that if undetectable corruption is allowed, then
the STP is not solvable.  Neither is it solvable if all messages can
be lost.  Thus, following~\cite{HZ}, we assume (a) that all
corruptions are detectable and (b) a strong fairness condition: for
any given link $l$, if infinitely often a message is sent on $l$,
then infinitely often some message is delivered on $l$.
We formalize strong fairness by restricting to systems where
$\FairS(l)$ holds for all links $l$.
The safety and liveness conditions for STP are run-based
specifications. As argued by Fagin et al.~\cite{FHMV94}, it is
often better to think in terms of knowledge-based specifications for
this problem.  The real goal of the STP is to get the receiver to
know the bits.
Writing $\ff{ K_R(X(n))}$ as an abbreviation for
$K_R(X(n) = 0) \lor K_R(X(n) = 1)$,
we really want to satisfy the knowledge-based specification
$$\ff{ \varphi^{kb}_{stp}\eqdef \forall n\: \diamondsuit K_R(X(n))}.$$
This is the specification we now synthesize.

Since we are assuming fairness, $S$ can ensure that $R$ learns the
$n$th bit by sending it sufficiently often. Thus, $\ff{S}$ can
ensure that $\ff{R}$ learns the $\ff{n}^{th}$ bit if, infinitely
often, either $\ff{ S}$ sends $\ff{ X(n)}$ or $\ff{S}$ knows that
$\ff{ R}$
knows $\ff{X(n)}$.
(Note that once $S$ knows that $R$ knows $X(n)$, $S$ will continue
to
know this, since  local states encode histories.)
We can enforce this by using an appropriate instantiation of ${\ff
Fair^{kb}}$.
Let $\boldc_S$ be a (nonrigid) constant that, intuitively, represents the
smallest $n$ such that $S$ does not know that $R$ knows $X(n)$, if such
an $n$ exists.
That is, we want the following formula to be true:
$$\exists n.~\neg K_S K_R X(n)) \rimp (\neg K_S K_R X(\boldc_S) \land
\forall k < \boldc_S.~K_S K_R X(k) ).$$
We abbreviate the formula $\forall k < n.~K_S K_R
(X(k)) \wedge \neg K_S K_R(X(n))$ as
$K_S K_R (X[0..n))$.

Let $\varphi_S$ be the knowledge-based formula that holds
at a consistent cut $c$
if
and only if
there exists a smallest $n$ such that, at $c$, $S$ does not know
that $R$ knows $X(n)$:
$$\ff {\varphi_S \eqdef \exists n. ~K_S K_R (X[} 0 .. n\ff{))}.$$
\commentout{ Let $f_S$ be the function that, given a value $n$ and a
local state
$s_S$ of $S$, returns the pair $(n,X(n))$;
$f_S$ is well defined, since $X$ is a local variable of $S$,
and is definable in Nuprl.
}%
Let $t_S$ be the term $\langle \boldc_S, X(\boldc_S) \rangle $.%
\footnote{We are implicitly assuming here that the pairing function that
maps $x$ and $y$ to $\langle x, y\rangle$ is in the language.}
Let $l_{SR}$ denote the communication link from $S$ to $R$. Now
consider the knowledge-based specification
${\ff Fair^{kb}_I(\phi_S, t_S,l_{SR})}$.
${\ff Fair^{kb}_I(\phi_S, t_S,l_{SR})}$ holds in a system
$\Sys$ if,
(1)
whenever $\ff R$ receives a message from $\ff S$, the
message is a pair of the
form $\langle n,X(n)\rangle$; (2) at the time $\ff S$ sent
this message to $\ff R$, $\ff S$ knew that $\ff R$ knew the first
$\ff n$ elements in the sequence $\ff X$, but $\ff S$ did not know
whether $\ff R$
knew $\ff X(n)$; and (3)
$\ff R$ is guaranteed to either eventually receive the
message $\langle n, X(n) \rangle$ or eventually know  $\ff X(n)$.

How does the sender learn which bits the receiver knows?  One
possibility is for  $\ff{ S}$ to receive
from $\ff{ R}$ a request to send $\ff{ X(n)}$. This can be taken by
$\ff{S}$ to be a signal that $R$ knows all the preceding bits.
We can ensure that $S$ gets this information by again using an
appropriate instantiation of $\ff{Fair^{kb}}$. Define
$\boldc_R$ be a (nonrigid) constant that, intuitively, represents the
smallest $n$ such that $R$ does not know $X(n)$, if such
an $n$ exists. In other words, we want the following formula to be true:
$$\exists n. ~\neg K_R X(n) \rimp (\neg K_R X(\boldc_R) \land \forall
k<\boldc_R. ~K_R X(k)).$$

We abbreviate $\forall k<n. ~ K_R(X(k)) \wedge \neg K_R(X(n))$ simply as
$K_R(X[0..n))$.
\commentout{
$t_R$ be a term defined as the smallest $n$ such that
that $R$ does not know $X(n)$, if such an $n$ exists, or $\infty$
otherwise.
}
We take $\varphi_R$ to be the knowledge-based formula
$$\ff {\varphi_R \eqdef \exists n.~K_R(X[} 0..n\ff{))},$$
which says that there exists a smallest $n$ such that $R$ does not
know $X(n)$ (or, equivalently, such that $\boldc_R =n$ holds).
\commentout{ Let $f_R$ be the function that,
given $R$'s local state,
 returns the least index
$n$ such that $R$
does not know $\ff{X(n)}$.
Since $R$'s local state determines what $R$ knows and does not know
$R$ can indeed compute $n$ from its local state.
Just like $f_S$, $f_R$ is definable in Nuprl.
}%
Finally, let $l_{RS}$ denote the communication link from $R$ to $S$.
$\ff Fair^{{\ff kb}}_I(\varphi_R, t_R,l_{RS})$ implies that
whenever $\ff S$ receives a message
$n$ from $R$, it is the case that, at the time $\ff R$ sent this
message, $\ff R$ knew the first $\ff n$ elements of $\ff X$, but not
$\ff X(n)$.
Note that, for all $n$, $\ff S$ is guaranteed to eventually receive
a  message $n$ unless $\ff R$ eventually knows $\ff X(n)$.

We can now use the Nuprl system to verify
our informal claim that we have refined the initial specification
$\ff{ \varphi^{kb}_{stp}}$. That is, the Nuprl system can prove
$$
\begin{array}{l}
\ff{(Fair^{kb}_I(\varphi_S, t_S,l_{SR})\wedge
Fair^{kb}_I(\varphi_R, \boldc_R,l_{RS}) \wedge } \\
~~~\ff{((\exists n.~\neg K_S K_R X(n)) \rimp K_S K_R X[}0..\ff{\boldc_S)) \wedge }\\
~~~\ff{((\exists n.~\neg K_R X(n)) \rimp K_R X[}0..\ff{\boldc_R)))
\Rightarrow \varphi^{kb}_{stp}}.
\end{array}
$$
No new techniques are needed for this proof: we simply unwind the
definitions of the semantics
of knowledge formulas and of the fairness specifications, and
proceed with
a standard proof by induction on the smallest $\ff n$ such that $\ff
R$  does not know $\ff X(n)$.

It follows from Lemma~\ref{lem:ex fair multiple k} that
$\ff{Fair^{kb}_I(\varphi_S, t_S,l_{SR})\wedge
Fair^{kb}_I(\varphi_R, \boldc_R,l_{RS})}$ is satisfied by the
combination of two simple knowledge-based programs, assuming that
message communication on links $\ff l_{SR}$ and $\ff l_{RS}$
satisfies the strong fairness conditions $\ff \FairS^{kb}(l_{SR})$ and
$\ff \FairS^{kb}(l_{RS})$. That is, for any two distinct actions
$\ff a_S$ and $\ff a_R$, the following is true:
$$
\begin{array}{l}
\ff{Fair\mbox{-}Pg(\varphi_S, t_S, l_{SR},a_S)\oplus
Fair\mbox{-}Pg(\varphi_R, \boldc_R, l_{RS}, a_R) \psatkbI}\\
\quad \ff{(\FairS^{kb}(l_{SR})\wedge \FairS^{kb}(l_{RS}))
\Rightarrow }
 \ff{(Fair^{kb}_I(\varphi_S, t_S,l_{SR})\wedge Fair^{kb}_I(\varphi_R,
 \boldc_R,l_{RS}))}.\\
\end{array}
$$
\commentout{ ${\bf f_S}$ and ${\bf f_R}$ are constant symbols, and
we define $I({\bf f_R}) = f_R$ and $I({\bf f_S}) = f_S$.
}%
As explained in Section~\ref{example: spec},
$\ff{\FairS^{kb}(l_{SR})\wedge \FairS^{kb}(l_{RS})}$ says that if
infinitely often a message is sent on $l_{SR}$ then infinitely often
a message is received on $l_{SR}$, and, similarly, if infinitely
often a message is sent on $l_{RS}$ then infinitely often  a message
is received on $l_{RS}$; as mentioned at the beginning of
this section, we restrict to systems where these conditions are met.
Furthermore, it is not difficult to show that we can use simple
initialization clauses to guarantee that the constraints on the
interpretation of $\boldc_S$ and $\boldc_R$ are satisfied:
$$
\begin{array}{l}
{\cf @}S~{\cf initially}~
\Box ((\exists n.~\neg K_S K_R X(n)) \rimp K_S K_R X[0..\boldc_S)) \psatkbI  \\
\quad \ff{(\exists n.~\neg K_S K_R X(n)) \rimp K_S K_R X[}0..\ff{\boldc_S)}, \\
\\
{\cf @}R~{\cf initially}~
\Box ((\exists n.~\neg K_R X(n)) \rimp K_R X[0..\boldc_R)) \psatkbI \\
\quad \ff{(\exists n.~\neg K_R X(n)) \rimp K_R X[}0..\ff{\boldc_R)}.
\end{array}
$$
Thus,
\commentout{
$$
\ff{Fair\mbox{-}Pg(\varphi_S, t_S, l_{SR},a_S)\oplus
Fair\mbox{-}Pg(\varphi_R, t_R, l_{RS},a_R) \psatkbI
\varphi^{kb}_{stp}}.
$$
}
$\ff{Pg^{kb}_S(\varphi_S, t_S, l_{SR},a_S)
\oplus
Pg^{kb}_R(\varphi_R, \boldc_R, l_{RS},a_R)) \psatkbI \varphi^{kb}_{stp}}$,
where
$$
\begin{array}{l}
\ff{Pg^{kb}_S(\varphi_S, t_S, l_{SR},a_S) } \eqdef
\ff{Fair\mbox{-}Pg(\varphi_S, t_S, l_{SR},a_S)} {\cf \oplus} \\
\quad \quad {\cf @}S~{\cf initially}~
\Box ( (\exists n.~\neg K_S K_R X(n)) \rimp K_S K_R X[0..\boldc_S)),\\
\\
\ff{{Pg}^{kb}_R(\varphi_R, \boldc_R, l_{RS},a_R)) } \eqdef
\ff{Fair\mbox{-}Pg}(\varphi_R, \boldc_R, l_{RS},a_R) {\cf \oplus }\\
\quad \quad {\cf @}R~{\cf initially}~
\Box ((\exists n.~\neg K_R X(n)) \rimp K_R X[0..\boldc_R)).
\end{array}
$$
From the definition of $\ff{Fair\mbox{-}Pg}(\varphi_R, \boldc_R,
l_{RS},a_R)$ in Section~\ref{example: spec}, it follows that
$\ff{Pg^{kb}_S}(\varphi_S, t_S,$ $l_{SR},$ $a_S)$ is the following
composition:
$$
\begin{array}{l}
{\cf @}S~{\cf initially}~
\Box ((\exists n.~\neg K_S K_R X(n)) \rimp K_S K_R X[0..\boldc_S)) {\cf \oplus} \\
{\cf @}S~\kind=\local(a_S)~  {\cf only~if}~ \ff{\exists n.~K_S K_R(X[}0..\ff{n))}~
    \cf{\oplus} \\
{\cf @} S~{\cf if}~\kind=\local(a_S) ~{\cf then ~msg(}{\ff
l_{SR}}{\cf )\, := \,}
t_S  ~\cf{ \oplus} \\
{\cf @}{\ff S}~{\cf only ~events ~in~ [}{\ff a_S}{\cf ] ~ affect
~msg(}{\ff l_{SR}}){\cf \oplus}\\
{\cf @}S~{\cf if~necessarily}~\ff{\exists n.~K_S K_R(X[}0..\ff{n))}~{\cf
then~i.o.}~\kind=local(a_S). \commentout{
{\cf @}S~{i.o.}~\kind=\local(a_S)~  {\cf if~i.o.}~
\ff{\exists n.~K_SK_R(X[}0..\ff{n)).}
}%
\end{array}
$$

\noindent Using the program notation of Fagin et al.~\cite{FHMV},
$\ff{Pg^{kb}_S(\varphi_S, t_S, l_{SR},a_S)}$
is
essentially
semantically equivalent to the following collection of programs, one
for each
value $n$:
$$ {\cf if}~ {\kf K_S (K_R X(0) \wedge \ldots \wedge
K_R X(n-1))\wedge \neg K_S K_R X(n)} ~{\cf then} ~{\ff
send}_{l_{SR}}(\langle n, X(n) \rangle) ~{\cf else}~{\ff skip}.$$
In both of these programs, $S$ takes the same action under the same
circumstances, and with the same effects on its local state.
That is,
given a run $r$ (i.e., a sequence of global states) consistent with
the collection of knowledge-based programs, we can construct an
event structure
$\es$ consistent with $\ff{Pg^{kb}_S(\varphi_S, t_S,
l_{SR},a_S)}$ such that the sequence of local states of $S$ in
$\es$, with stuttering eliminated,  is the same as in $r$. The
converse is also true. More precisely, in a run $r$ consistent with
the
collection of knoweldge-based programs, at each point of time,
either $\ff S$ knows that $\ff
R$ knows the value of $X(n)$ for all $n$, or there exists
a smallest $\ff n$ such that $\neg K_S K_R(X(n))$ holds.  In the
first case, $S$ does nothing, while in
the second case $S$ sends $\langle n, X(n)\rangle$ on $\ff l_{SR}$.
Similarly, in an event structure $\es$ consistent with
$\ff{Pg^{kb}_S(\varphi_S, t_S,  l_{SR},a_S)}$,
if $\ff S$ knows that $\ff R$ knows $\ff X(n)$ for all $n$, then
$\ff S$ does nothing; if not, then
it is impossible for $\ff S$ to know that $\ff R$ knows the first
$n$ bits, but never know that $\ff R$ knows $\ff
X(n)$, without eventually
$\ff S$ taking an $\ff a_S$  action with value $\ff \langle n,
X(n)\rangle$. This means that for each run $\ff r$ consistent with
the collection of knowledge-based programs, the event structure
$\es$ in which $\ff S$ starts from the same initial state as in
$\ff r$  and performs action $\ff a_S$ as soon as it is enabled has
the same sequence of local states of $\ff S$ as $\ff r$. For each
event structure $\es$ consistent with
$\ff{Pg^{kb}_S(\varphi_S, t_S,  l_{SR},a_S)}$, in the run $\ff
r$ of global states in
$\es$ with stuttering eliminated, $\ff S$ takes action $\ff a_S$ as
soon as enabled; subsequently,  $r$ is consistent with the
collection of knowledge-based programs.

Similarly,
$\ff{Pg^{kb}_R(\varphi_R, \boldc_R,  l_{RS},a_R)}$ is
essentially
semantically equivalent to
the following collection of programs, one for each value $n$:
$$ {\cf if}~ {\kf  K_RX(0)\wedge \ldots \dots \wedge
K_R X(n-1)\wedge \neg K_R X(n)}~ {\cf then} ~{\ff
send}_{l_{RS}}(n)~{\cf else}~{\ff skip}.$$ Thus, the derived program
is essentially one of the knowledge-based programs considered by
Halpern and Zuck~\cite{HZ}. This is not surprising, since our
derivation followed much the same reasoning as that of Halpern and
Zuck. However, note that we did not first give a  knowledge-based
program and then verify that it satisfied the specification. Rather,
we derived the knowledge-based programs for the sender and receiver
from the proof that the specification was satisfiable.  And, while
Nuprl required ``hints''
in terms of what to prove, the key ingredients of the proof, namely,
the specification $\ff{Fair^{kb}_I(\varphi, t ,l)}$ and the proof
that $\ff{ Fair\mbox{-}Pg(\varphi, t, l,a)}$ realizes it, were
already in the system, having been used in other contexts. Thus,
this suggests that we may be able to apply similar techniques to
derive programs satisfying other specifications in communication
systems with only weak fairness guarantees.

\subsection{Synthesis of standard programs for STP}
This takes care of the first stage of the synthesis process.  We now
want to find a standard program that implements the knowledge-based
program.  As discussed by Halpern and Zuck \cite{HZ}, the exact
standard program that we use depends on the underlying assumptions
about the communications systems.  Here we sketch an approach to
finding such a standard program.

The first step is to identify the exact properties of knowledge that
are needed for the proof.
This can be done by inspecting the proof to see
which properties of the knowledge operators $\ff{ K_S}$ and $\ff{
K_R}$ are used.
The idea is then to replace formulas involving the knowledge operators
by standard (non-epistemic formulas) which have the relevant
properties.

\commentout{ We  can then replace $\ff{ \past (K_R (X(i)=v))}$ by an
abstract predicate $\ff{ Q (X(i)=v)}$ and $\ff{ K_S (\past K_R
(X\left[ i\right]=v))}$ by $\ff{ P(X(i)=v)}$. As before, we
abbreviate $\ff{ P(X(i)=0)\vee P(X(i)=1)}$ as $\ff{ P(X(i)}$, and
$\ff{ Q(X(i)=0) \vee }$ $\ff{Q(X(i)=1)}$ as $\ff{ Q(X(i)}$. We add
as hypotheses all the identified properties, now as properties of
$\ff{ Q}$ and $\ff{ P}$, and check whether the former proof still
applies. If not, we add whatever additional properties are needed.
Note that we can use Nuprl to automate these checks.

This approach enables us to prove that the specification is
satisfiable in a more general setting. The specification is now
written in terms of $\ff{ P}$ and $\ff{ Q}$, and denoted
$\ff{\tilde{\varphi}^{kb}}$. $\ff{ Fair^{kb}(P_R^{kb}, i,l_{RS})}$
is replaced by $\ff{ Fair^{kb}}(\tilde{Q}, i,l_{RS})$ with
$$\ff{ \tilde{Q}(i)\stackrel{def}{\equiv} \forall j<i \: Q(X(j)) \wedge
\neg Q(X(i))};$$ similarly, $\ff{ Fair^{kb}}\ff{ (P_S^{kb}, \langle
i,X\left[i\right]\rangle,l_{SR})}$ becomes $\ff{ Fair^{kb}}\ff{
(\tilde{P}, \langle i, X\left[i\right]\rangle,l_{SR})}$ with $$\ff{
\tilde{P}(i)\stackrel{def}{\equiv} \forall j<i\: P(X(i)) \wedge \neg
P(X(i))}.$$ }
Suppose that $\tilde{\phi}_S^{kb}$ is a formula
that has a free variable $\idx$, and is
guaranteed to be an $S$-formula in all interpretations $I$.
Roughly speaking, we can think of $\tilde{\phi}_S^{kb}$ as
corresponding to $\ff K_S K_R(X(\idx))$.

Let $\phi_S^{kb}$ be an abbreviation of
$${\ff \exists n. ~((\forall k<n.~ \tilde{\phi}_S^{kb}[ \idx / k])\wedge \neg
\tilde{\phi}_S^{kb}[\idx / n])}.$$
Thus, $\phi_S^{kb}$ is the analogue of $\phi_S$ in
Section~\ref{sec:kbstp}.
\commentout{
Similarly, let $P_R^{kb}$ be a predicate on systems, natural
numbers, and local states of
agent $\ff R$, and let $\phi_R^{kb}$ be an abbreviation of
$${\ff \exists n. ~((\forall k<n.~ P_R^{kb}(Sys,k,s_R))\wedge \neg
P_R^{kb}(Sys,n,s_R))}.$$
}
Similarly, suppose that $\tilde{\phi}_R^{kb}$ is a formula that
has a free variable $\idx$, and is
guaranteed to be an $R$-formula in all interpretations $I$;
let $\phi_R^{kb}$ be an abbreviation of
$${\ff \exists n. ~((\forall k<n.~ \tilde{\phi}_R^{kb}[\idx / k])\wedge \neg
\tilde{\phi}_R^{kb}[\idx / n])}.$$
We can think of $\tilde{\phi}_R^{kb}$ as corresponding to
$K_R(X(\idx))$.
\commentout{
In the following, we assume that ${\cf P_S^{kb}}$, ${\cf P_R^{kb}}$,
${\cf {\phi}_S^{kb}}$, and ${\cf {\phi}_R^{kb}}$ are predicate
symbols such that $I'({\cf P_S^{kb}})=P_S^{kb}$, $I'({\cf
P_R^{kb}})=P_R^{kb}$, $I'({\cf {\phi_S}^{kb}})={\phi}^{kb}$, and
$I'({\cf {\phi_R}^{kb}})={\phi_R}^{kb}$.}
We also use
constants $\tilde{\boldc}_S$ and $\tilde{\boldc}_R$ that are
analogues of
$\boldc_S$, $\boldc_R$;
$\tilde{\phi}_S^{kb}$
plays the same role in the definition of $\tilde{\boldc}_S$ as $K_S
K_R(X(\idx))$ played in the definition of $\boldc_S$, and
$\tilde{\phi}_R^{kb}$ plays the same role in the definition of
$\tilde{\boldc}_R$ as $K_R(X(\idx))$ played in the definition of $\boldc_R$.
Thus, we take $\tilde{\boldc}_S$ to be a constant that represents the least $n$
such that $\tilde{\phi}_S^{kb}[\idx / n]$ does not hold
(that is, we want $(\exists n.~\neg \tilde{\phi}_S^{kb}[\idx / n ]) \rimp (\forall k<\boldc_S. ~\tilde{\phi}_S^{kb}[\idx / k ] \wedge \neg \tilde{\phi}_S^{kb}[\idx / \boldc_S ])$ to be true),
and define $\tilde{t}_S$ as the pair $\langle \tilde{\boldc_S},
X(\tilde{\boldc_S})\rangle$.
Similarly, we take $\tilde{\boldc}_R$ to be a constant that represents
the least $n$
such that $\tilde{\phi}_R^{kb}[\idx / n ]$ does not hold
(that is, we want $(\exists n. ~\neg \tilde{\phi}_R^{kb}[\idx / n]) \rimp (\forall k < \boldc_R.~\tilde{\phi}_R^{kb}[\idx / k] \wedge \neg \tilde{\phi}_R^{kb}[\idx / \boldc_R])$
to be true).
\commentout{
We now replace $I'(\phi_S)$, $I'(\phi_R)$, $t_S$, and $t_R$ in
$\ff{Fair^{kb}(I'(\varphi_S),I'(t_S),l_{SR})}$ and ${\ff
Fair^{kb}(I'(\varphi_R),I'(t_R),l_{RS})}$  with ${\phi}_S^{kb}$,
${\phi}_R^{kb}$,
$\hat{t}_S$, and $\hat{t}_R$, respectively.
}

Let $\ff{\tilde{\varphi}_\mathit{stp}^{kb}(\tilde{\phi}_R^{kb})}$ be the
specification
that results by using $\tilde{\phi}_R^{kb}[\idx/n]$ instead of $K_R(X(n))$ in
$\varphi^{kb}_{\mathit{stp}}$:
$$\ff{\tilde{\varphi}_{\mathit{stp}}^{kb}(\tilde{\phi}_R^{kb})}\eqdef {\ff
\forall n.~\diamondsuit \tilde{\phi}_R^{kb}[\idx / n]}.$$
We prove the goal $\tilde{\varphi}_{\mathit{stp}}^{kb}(\tilde{\phi}_R^{kb})$ by refinement:
at each step, a rule (or tactic) of Nuprl is applied, and a number of
subgoals (typically easier to prove) are generated; the rule gives a
mechanism of constructing a proof of the goal from proofs of the
subgoals. Some of the subgoals cannot be further refined in an
obvious manner; this is the case,
for example,
for the simple conditions on $\ff {\tilde{\phi}_S^{kb}}$ or $\ff {\tilde{\phi}_R^{kb}}$.
The new theorem states that, under suitable
conditions on $\ff{\tilde{\phi}_S^{kb}}$ and $\ff{ \tilde{\phi}_R^{kb}}$,
$\ff{{\varphi}^{kb}(\tilde{\phi}_R^{kb})}$ is satisfiable if both
$\ff{Fair}_I^{kb} (\phi_S^{kb}, \tilde{t}_S , l_{SR})$ and $
\ff{Fair}_I^{kb} (\phi_R^{kb}, \tilde{\boldc}_R , l_{RS})$ are
satisfiable.%
\footnote{The Nuprl lemma that corresponds to this result can be viewed at
\begin{center}
http://www.cs.cornell.edu/info/projects/nuprl/fdlcontent/%
p0\_963683\_/send-minimal-realizable.html.
\end{center}
  For ease of exposition, we have simplified and modified some Nuprl
notation in our presentation in this paper.  The
differences between the Nuprl lemma and the result of the paper are
discussed at http://www.cs.cornell.edu/home/halpern/papers/synthesis-appendix.pdf.}

We now explain the
conditions placed on the predicates $\ff {\tilde{\phi}_S^{kb}}$
and  $\ff {\tilde{\phi}_R^{kb}}$.
One condition is that $\ff {\tilde{\phi}_R^{kb}}$ be {\em stable}, that is, once
true, it stays true:
\commentout{
$$
\begin{array}{ll}
\ff{\Stable(P_R^{kb})} \eqdef & ~\ff{\lambda \Sys.~\forall \es\in
\Sys.~\forall e_R@R\in \es.~
  \forall n.~ P_R^{kb}(\Sys,n,\state~\before~e_R) \Rightarrow }\\
& \quad \ff{P_R^{kb}(\Sys,n,\state~\after~e_R)}.\\
\end{array}
$$
Assuming $\ff{\Stable(P_R^{kb})}$ allows us to prove
$\ff{{\varphi}^{kb}(P_R^{kb})}$ by induction on the least index $\ff
n$ such that $\ff \neg P_R^{kb}(\Sys, $ $\ff n, \state)$ holds.
}
$$
\begin{array}{ll}
\ff{\Stable(\tilde{\phi}_R^{kb})} \eqdef &\!\!\!\lambda \Sys.~\forall \es\in
\Sys.~\forall e_R@R\in \es.~
\forall n. ~ I(\tilde{\phi}_R^{kb}[\idx/n])(\Sys,\state~\before~e_R)
\Rightarrow \\
& \quad \ff{I(\tilde{\phi}_R^{kb}[\idx/n])(\Sys,\state~\after~e_R)}.
\end{array}
$$
Assuming $\ff{\Stable(\tilde{\phi}_R^{kb})}$ allows us to prove
$\ff{{\varphi}_R^{kb}}$ by induction on the least index $\ff
n$ such that $\ff {\neg \tilde{\phi}_R^{kb}[\idx / n]}$ holds.

To allow us to carry out a case analysis on whether
$\ff {\tilde{\phi}_R^{kb}}$ holds,
we
also assume that $\ff {\tilde{\phi}_R^{kb}}$ satisfies the principle
of excluded middle; that is, we assume that
$\ff{\Dec(\tilde{\phi}_R^{kb}) \eqdef }$ $\ff{\Dec(\forall n. }$ $\ff{(\tilde{\phi}_R^{kb}[\idx / n])^t)}$.
For similar reasons, we also restrict $\ff {\tilde{\phi}_S^{kb}}$ to being stable
and determinate; that is, we require that $\ff{\Stable(\tilde{\phi}_S^{kb})} $
and $\ff \Dec(\tilde{\phi}_S^{kb})$ both hold.

The third
\commentout{
condition we impose establishes a connection between $\ff P_S^{kb}$
and $\ff P_R^{kb}$, and ensures that, if $\ff P_S^{kb}$ holds for
some value $\ff n$, then eventually $\ff P_R^{kb}$ will also hold
for $\ff n$:
$$
\begin{array}{l}
\ff{\Imp(P_S^{kb},P_R^{kb})} \eqdef \\
\quad  \ff{\lambda \Sys. ~\forall
\es\in \Sys.~\forall n.~\forall e_S@S\in \es.~
  P_S^{kb}(\Sys, n,\state~\before~e_S)\Rightarrow}\\
\quad ~~ \ff{\quad \exists e_R\succ e_S@R\in \es.~ P_R^{kb}(\Sys, n,\state~\after~e_R)}.\\
\end{array}
$$
}
condition we impose establishes a connection between $\ff
{\tilde{\phi}_S^{kb}}$
and $\ff {\tilde{\phi}_R^{kb}}$, and ensures that, for all values $n$, if
$\ff {\tilde{\phi}_S^{kb}[ \idx / n ]}$ holds, then eventually $\ff {\tilde{\phi}_R^{kb}[\idx/n]}$ will also hold:
$$
\begin{array}{l}
\ff{\Imp(\tilde{\phi}_S^{kb},\tilde{\phi}_R^{kb})} \eqdef \\
\quad  \lambda \Sys. ~\forall
\es\in \Sys.~\forall n.~\forall e_S@S\in \es.~  I
(\tilde{\phi}_S^{kb}[\idx/n])(\Sys,\state~\before~e_S)\Rightarrow\\
\quad  \ff{\quad \exists e_R\succ e_S@R\in \es.~ I
(\tilde{\phi}_R^{kb}[\idx/n])(\Sys,\state~\after~e_R)}.\\
\end{array}
$$

\noindent To explain the next condition,
recall that $\ff{\tilde{\phi}_R}$ is meant to represent $K_R(X(\idx))$.
With this interpretation, $
I(\forall k\le
n.~\tilde{\phi}_R^{kb[m/k]})(\Sys,\state~\before~\send(e_S))$
says that $\ff R$ knows the first $\ff n$ bits before it sends a
message to $\ff S$.
We would like it to be the case that, just as with the
knowledge-based derivation,
when $S$ receives $R$'s message, $\ff S$ knows that $\ff R$ knows
the  $\ff n^{th}$ bit.
Since we think of $\ff {\tilde{\phi}_S^{kb}}$ as saying that
$K_SK_R(X(\idx))$ holds,
we expect $I(\tilde{\phi}_S^{kb}[\idx/n])(\Sys,\state~\after~e_S)$ to
be true.
Define $\ff{\Rcv(\tilde{\phi}_S^{kb}, \tilde{\phi}_R^{kb}, l_{RS})}$ to be an abbreviation of
$$
\begin{array}{l}
\ff{\lambda \Sys. \forall \es\in \Sys.~\forall e_S@S\in \es.~
(\kind(e_S)=\rcv(l_{RS}))\Rightarrow }\\
\quad \begin{array}{ll}
      \ff{\forall n.~} & \ff{ (\forall k\le n.~
    I(\tilde{\phi}_R^{kb}[\idx/n])(\Sys,\state~\before~\send(e_S)))\Rightarrow}\\
     & \ff{    I(\tilde{\phi}_S^{kb}[\idx/n])(\Sys,\state~\after~e_S)}.\\
    \end{array}
\end{array}
$$
With this background, we can describe the last
condition. Intuitively, it says that if $\ff n$ is the least value
for which $\ff {\tilde{\phi}_S^{kb}}$ fails when $\ff S$ sends a message to $\ff
R$, then $\ff {\tilde{\phi}_R^{kb}}$ holds for $\ff n$ upon message delivery:
$$
\begin{array}{l}
\ff{\Rcv(\tilde{\phi}_R^{kb}, \tilde{\phi}_S^{kb}, l_{SR}) \tbar}~\\
\quad
\ff{\lambda \Sys.~\forall \es\in \Sys.~~\forall e_R@R\in \es. ~(\kind(e_R)=\rcv(l_{SR}))\Rightarrow }\\
  \quad \ff{\forall n.~
(\forall k < n.~I(\tilde{\phi}_S^{kb}[m /k])(\Sys,\state~\before~\send(e_R)) \wedge }\\
\quad \quad \quad \ff{
I(\neg \tilde{\phi}_S^{kb}[m /n])(\Sys,\state~\before~\send(e_R))}\\
  \quad \quad \quad \ff{
\rimp I(\tilde{\phi}_R^{kb}[\idx/n])(\Sys,\state~\after~e_R))}.
\end{array}
$$

\commentout{ One condition requires that $\ff P^{kb}$ and $\ff
Q^{kb}$ be {\em stable}, i.e., once true, they stay true. This is
clearly true if we think of  $\ff P^{kb}$ and $\ff Q^{kb}$ as
interpretations of $\ff K_S K_R (X(i))$ and $\ff K_R(X(i))$,
respectively, since we have assumed that agents have perfect recall.
It is also required that $\ff P^{kb}$ and $\ff Q^{kb}$ satisfy the
principle of excluded middle; this allows us to refine goals by
proceeding with a case analysis on whether $\ff P^{kb}$ (resp. $\ff
Q^{kb}$) holds of a certain index and
local
state of $\ff S$ (resp. $\ff R$). Other conditions refer to the
transfer of information from $\ff S$ to $\ff R$; for example, we
require that, whenever $\ff P^{kb}$ holds of some index $\ff i$ when
$\ff S$ sends a message to $\ff R$, $\ff Q^{kb}$ holds of $\ff i$,
upon $\ff R$ receiving the message. }
\noindent We abbreviate the conjunction of
these
conditions as
$\ff{\psi^{kb}(\tilde{\phi}_S^{kb},\tilde{\phi}_R^{kb},\tilde{t}_S,\tilde{t}_R,
l_{SR},l_{RS})}$. The new theorem says
$$
\begin{array}{l}
\commentout{
\ff{\psi^{kb}(P_S^{kb}, P_R^{kb}, I'(\hat{t}_S), I'(\hat{t}_R),
l_{SR}, l_{RS}) \wedge Fair^{kb}({P_S}^{kb}, I'(\hat{t}_S),
l_{SR})\wedge
Fair^{kb}({P_R}^{kb},I'(\hat{t}_R), l_{SR}) }\\
\quad \quad {\ff \Rightarrow \varphi^{kb}(P_R^{kb}).}
}
\ff{\psi^{kb}(\tilde{\phi}_S^{kb}, \tilde{\phi}_R^{kb}, \tilde{t}_S,
\tilde{\boldc}_R,
l_{SR}, l_{RS}) \wedge } \\
\ff{Fair_I^{kb}({\phi}_S^{kb}, \tilde{t}_S,l_{SR})\wedge
Fair_I^{kb}({\phi}_R^{kb}, \tilde{\boldc}_R, l_{RS}) } \wedge  \\
\ff{( (\exists n.~\neg \tilde{\phi}_S^{kb}[\idx / n]) \rimp (\forall k<n.~\tilde{\phi}_S^{kb}[\idx / k] \wedge \neg \tilde{\phi}_S^{kb}[\idx / \boldc_S]))} \wedge \\
\ff{((\exists n.~\neg \tilde{\phi}_R^{kb}[\idx / n]) \rimp (\forall k<n.~\tilde{\phi}_R^{kb}[\idx / k] \wedge \neg \tilde{\phi}_R^{kb}[\idx / \boldc_R]))}\\
\quad \quad {\ff \rimp \varphi^{kb}(\tilde{\phi}_R^{kb}).}
\end{array}
$$

\noindent We can prove that the following is true for any two distinct actions
$\ff a_S$ and $\ff a_R$:
\commentout{
$$
\begin{array}{l}
\ff{(Fair\mbox{-}Pg}({\cf {P_S}^{kb}}{\ff  , t_S, l_{SR},a_S)\oplus
       Fair\mbox{-}Pg}({\cf {P_R}^{kb}}{\ff , t_R, l_{RS},a_R))  \psatkbI}\\
\quad \ff{(\psi^{kb}(P_S^{kb},P_R^{kb}, f_S, f_R,l_{SR},l_{RS})
\wedge \FairS(l_{SR}) \wedge \FairS(l_{RS})\Rightarrow \varphi^{kb}(P_R^{kb}))}.\\
\end{array}
$$
}
$$
\begin{array}{l}
\Pg_S^{kb}({\phi}_S^{kb}, \tilde{t}_S, l_{SR}, a_S) \oplus
\Pg_R^{kb}({\phi}_R^{kb}, \tilde{\boldc}_R, l_{RS}, a_R) \psatkbI  \\
\quad \psi^{kb}(\tilde{\phi}_S^{kb}, \tilde{\phi}_R^{kb},
\tilde{t}_S, \tilde{\boldc}_R,
l_{SR}, l_{RS}) \wedge \FairS(l_{RS})\Rightarrow
\varphi^{kb}_{\mathit{stp}}(\tilde{\phi}_R^{kb}),
\end{array}
$$
where
$$
\begin{array}{l}
\ff{Pg^{kb}_S({\phi}_S^{kb}, \tilde{t}_S, l_{SR},a_S) } \eqdef \\
\quad \ff{Fair\mbox{-}Pg({\phi}_S^{kb}, \tilde{t}_S, l_{SR},a_S)} {\cf \oplus} \\
\quad {\cf @}S~{\cf initially}~
\Box ((\exists n.~\neg \tilde{\phi}_S^{kb}[\idx / n]) \rimp (\forall k<\boldc_S.~\tilde{\phi}_S^{kb}[\idx / k] \wedge \neg \tilde{\phi}_S^{kb}[\idx / \boldc_S])),\\
\\
\ff{{Pg}^{kb}_R({\phi}_R^{kb}, \tilde{t}_R, l_{RS},a_R)) } \eqdef \\
\quad \ff{Fair\mbox{-}Pg(\phi_R^{kb}, \tilde{t}_R, l_{RS},a_R)} {\cf
\oplus }\\
\quad {\cf @}R~{\cf initially}~
\Box ((\exists n.~\neg \tilde{\phi}_R^{kb}[\idx / n]) \rimp (\forall k<\boldc_R.~\tilde{\phi}_R^{kb}[\idx / k] \wedge \neg \tilde{\phi}_R^{kb}[\idx / \boldc_R])).
\end{array}
$$
In particular, for the terms $t_S$ and $\boldc_R$ and
formulas $\ff \varphi_S$ and $\ff \varphi_R$
defined in the previous section,
we can show that $\psi^{kb}(\varphi_S^{kb},\varphi_R^{kb}, t_S,
\boldc_R,l_{SR},l_{RS})$ is true.
Thus, the new theorem is indeed a generalization of the previous
results.

The formulas $\ff{\phi_S}$ and $\ff{\phi_R}$ are not the only ones
that satisfy these conditions. Most importantly for the purpose of
extracting standard programs, the
conditions are satisfied by non-epistemic formulas, that is, formulas whose interpretations
do not depend on the entire system, just on the local states of the sender or the receiver agents,
respectively.
\commentout{
If, furthermore, the standard predicates satisfy the principle of
excluded middle,
}%
Note that
Lemma~\ref{lem: pg fair plus} guarantees
that the extracted program is \pgrealizable.
\commentout{
For example, we can choose $\ff P_R^{st}$ such that $\ff
P_R^{st}(i,s_R)$ is true exactly when $\ff s_R$ records that $\ff
R$ has received a message containing index $\ff i$, and choose $\ff
P_S^{st}$ such that $\ff P_S^{st}(i,s_S)$ is true if and only if
$\ff s_S$ records that $\ff S$ has received an index strictly
greater than $\ff i$.
}
\subsubsection{Stenning's protocol}
In the next two sections, we show
that by making relatively straightforward
choices for the formulas $\tilde{\phi}_S^{kb}$ and $\tilde{\phi}_R^{kb}$
and terms $\tilde{t}_S$ and $\tilde{t}_R$, we can derive
two well-known solutions for STP,
Stenning's protocol \cite{St} and an infinite-state variant of the
alternating-bit protocol~\cite{BSW}.
We start with Stenning's protocol.

In Stenning's protocol, the sender transmit the bits on
the tape in order
to the receiver.
The sender $S$ keeps track of the position $i_S$ of the bit in the
sequence that he will next
send to $R$, while the receiver $R$ keeps track of the first position
$i_R$ in the sequence for which he has not received the corresponding bit.
Initially, both $i_S$ and $i_R$ are set to $0$.
$S$ always sends $R$ message of the form $\langle X(i_S),i_S\rangle$.
When $R$ receives a message from $S$ whose second component is $i_R$, then $R$
increments $i_R$ and acknowledges the messages by sending $S$ the
message $i_R$;  $R$ disregards other messages.  If $S$
receives $i_R$ and $i_R > i_S$ (it is easy to see that this can happen
only if $i_R = i_S +1$), then $S$ increments $i_S$; $S$
disregards all other messages.
Note that it is straightforward to write clauses that
ensure that $i_S$ and $i_R$ indeed have these properties.
\commentout{
The following clauses do the job.
$$
\begin{array}{l}
\ff{{\cf @} S ~{\cf initially}~ i_S ~=~ 0 }~\oplus \\
\ff{{\cf @}S ~{\cf if}~ kind=local(inc_S)~ {\cf then}~ i_s := msg(l_RS)} ~\oplus \\
\ff{{\cf @}S ~kind=local(inc_S)~\cf{ only~ if~} ~ msg(l_RS)>i_S}~\oplus \\
\ff{{\cf @}S ~\cf{if ~necessarily~} msg(l_RS)>i_S ~ \cf{then ~i.o. ~} kind=local(inc_S)} ~\oplus \\
\ff{{\cf @}S ~{\cf only ~events~ in ~[}rcv(l_{RS}) {\cf ] ~affect~} i_S} ~\oplus ~\\
\ff{{\cf @}R ~{\cf initially } ~i_R ~= ~0 }~ \oplus\\
\ff{{\cf @}R ~{\cf if}~ kind=local(inc_R)~ {\cf then}~ i_R := i_R + 1} ~\oplus \\
\ff{{\cf @}R ~kind=local(inc_R)~\cf{ only~ if~} ~ p_2( msg(l_{SR})) = i_R}~\oplus \\
\ff{{\cf @}R ~\cf{if ~necessarily~}  p_2( msg(l_{SR})) = i_R~ \cf{then ~i.o. ~} kind=local(inc_R)} ~\oplus \\
\ff{{\cf @}R ~{\cf only ~events~ in ~[}rcv(l_{SR}) {\cf ] ~affect~} i_R},\\
\end{array}
$$
where
$p_2(a)$ represents the second element of a pair $a$.
}%
The clauses should say that initially both $i_S$ and $i_R$ are set to $0$, that $i_S$ only changes
when $S$ receives from $R$ a message larger than $i_S$, and that, if infinitely often this is the case,
then infinitely often $i_S$ is incremented; similarly, the clauses should say that $i_R$ only changes when
$R$ receives from $S$ a message whose last component is $i_R$, and if infinitely often this is the case, then
infinitely often $i_R$ is increased. As apparent from this short description, all such clauses can be expressed
in the message automata framework.
We can choose $\ff {{\tilde {\phi}}_R^{kb}}$ such that
$\ff{{\tilde \phi}_R^{kb}}(m)$
holds in $R$'s local state $\ff s_R$
exactly when $\ff s_R$ records that $\ff R$ has received a message
containing index $\ff \idx$
(that is, $\ff{{\tilde \phi}_R^{kb}}(m) \eqdef i_R > m$),
and choose $\ff {{\tilde \phi}_S^{kb}}$ such that $\ff {{\tilde {\phi}}_S^{kb}}(m)$ holds in
$S$'s local state $\ff s_S$
exactly when $\ff s_S$ records that $\ff S$ has received an index strictly
greater than $\ff \idx$
(that is, $\ff {{\tilde \phi}_S^{kb}} \eqdef i_S> m$)
.
It is not difficult to show that
\commentout{
$\ff
{\psi(P_S^{st},P_R^{st},I(\hat{t}_S),I(\hat{t}_R),l_{SR},l_{RS})}$
holds, where
$\ff \psi$ is just like $\ff \psi^{kb}$, except that now the
conditions in $\ff \psi^{kb}$   are not knowledge-based.
}
$\ff{\psi^{kb}({\tilde \phi}_S^{kb}, {\tilde \phi}_R^{kb}, \tilde{t}_S, \tilde{\boldc}_R,
l_{SR},l_{RS})}$ holds,
except that now this specification is not knowledge-based.
Note that $\ff \phi_S^{kb} (= \exists n.~ (\forall k<n.~ {\tilde{\phi}}_S^{kb}[ m/k])
\wedge \neg {\tilde{\phi}}_S^{kb}[m/n]) = \true$ and, similarly, $\ff \phi_R^{kb} = \true$.
In addition, $\ff \varphi^{kb}_{stp}(\tilde{\varphi}^{kb}_R)$ implies $\ff
\varphi^{kb}_{stp}$, which means that, assuming message
communication is fair,
$$\ff{Pg_S(\phi_S^{kb},\tilde{t}_S, l_{SR},a_S)\oplus Pg_R(\phi_R^{kb}, \tilde{t}_R, l_{RS},a_R)}$$
(together with the basic clauses ensuring that the
variables $i_S$ and $i_R$ behave appropriately)
satisfies the STP specification,
as long as $a_S$ and $a_R$ are distinct actions.
Note that the
program
$\ff{Pg_S(\phi_S^{kb}, \tilde{t}_S, l_{SR},a_S)\oplus }$
$\ff{Pg_R(\phi_R^{kb}, \tilde{t}_R ,l_{RS},a_R)}$
is realizable.
We have thus extracted a program that realizes the STP
specification.
Moreover, we can show that this program is essentially semantically
equivalent to Stenning's protocol.

\commentout{ For example,
suppose that $\ff{ S}$ has a
local
state variable $\ff{ i_S}$ such that
$\ff{  PX[0\dots i_S)}$ holds at the current
local
state of $\ff{ S}$; similarly, $\ff{ R}$ has a
local
state variable $\ff{ i_R}$ such that
$\ff{ Q X[0\dots i_R)}$ holds.
We can simply define
$$\ff{ P'(X(k))\stackrel{def}{\equiv} i_S\geq k} ~\mbox{and}~\ff{ Q'(X(k))\stackrel{def}{\equiv} i_R\geq k};$$
the resulting program is exactly  Stenning's \cite{St} protocol.
}

The Nuprl system is semi-automatic, in the sense that the
programmer indicates at
each step which refinement rule to apply. Users can group a sequence of
rules together into what is called a tactic.
In the discussion above, we did not apply any Nurpl tactics in the
derivation.  However, the reader can easily check that each refinement
step in the
proof outlined above
is either a basic refinement rule (i.e., induction,
case analysis for a formula satisfying the principle of excluded middle),
or an instance of the fairness specification from Section~\ref{example:
spec}.

The key point here is that by replacing the knowledge tests by
stronger predicates that imply them and do not explicitly mention
knowledge, we can derive standard programs that implement the
knowledge-based program. We believe that other standard
implementations of the knowledge-based program can be derived in a
similar way, although we have not yet carried out the derivation.

\subsubsection{The alternating-bit protocol}
Stenning's protocol works even if messages can be drop\-ped or duplicated,
and messages can be reordered.  All that is required is that
communication is fair, in the sense that a message sent infinitely often
is eventually received.  The alternating-bit protocol also works in an
environment where messages can be dropped or duplicated, but it does
require that messages are received in the order in which they are sent.
The advantage of making this extra assumption is that now a finite-state
protocol can be used.  Instead of using counters $i_S$ and $i_R$ to keep
track of which prefix of the sequence has been received, it suffices to
use a bit that alternates in value to do this.

In more detail, the sender
starts by reading the first value, stores it in the variable $x_S$, and sends
$(x_S,i_S)$ to the
receiver, where $i_S$ is a bit initially set to $0$. The
receiver maintains a bit $i_R$, initialized to $\lambda$ (a
null value). Upon receiving a message $(x_S,i_S)$ from the sender,
if $i_S \ne i_R$, then the receiver sets $i_R= i_S$, writes
$x_S$, and acknowledges $(x_S,i_S)$ (by sending $i_R$ to the sender);
if $i_S = i_R$, then the receiver ignores the message.
When the sender receives a message $i_R$ from the receiver with
$i_R = i_S$, then the sender reads the next bit in the sequence into
$x_S$ and sets $i_S$ to $1 - i_S$; otherwise, the sender ignores the message.
(Note that the values of $i_S$ alternates between 0 and 1, hence
the name of the protocol.)
\commentout{
To model this, we assume that agent $S$ has three local variables: a bit
$i_S$,
initially set to $0$,
\commentout{
a variable $j$ corresponding to the index in the
array $X$ the sender
is currently sending to the receiver, initially set to $0$,%
\footnote{In practice, the array is sent to $S$ externally, and the
variable $j$ is not needed, so that the alternating-bit protocol is a
finite-state protocol.}
}
The correctness shouldn't depend on it.
and a
variable $\ff{cnt}_S$
indicating how many times $i_S$ has been flipped so far.
The assumption is that, whenever $S$ flips the bit $i_S$, $x_S$ takes
the next value on the tape $X$. (In other words, $x_S = X(\ff{cnt}_S)$ holds.)
We also assume
that the receiver $R$
has two local variables: $i_R$, taking values $\lambda$,
$0$, or $1$,
initially set to $\lambda$, and $\ff{cnt}_R$, which indicates how many
times $i_R$ has been flipped so far.
}
Let ${cnt}_S$ be a variable representing how many times the bit $i_S$ has been flipped;
similarly, let ${cnt}_R$ be a variable representing how many times the bit $i_R$ has been flipped.
Let $\tilde{\phi}_S^{kb} \eqdef ({cnt}_S \ge m+1)$, $\tilde{\phi}_R^{kb} \eqdef ({cnt}_R \ge m+1)$,
$\tilde{t}_S \eqdef \langle x_S, i_S\rangle$, and $\tilde{t}_R \eqdef i_R$.
Intuitively, whenever ${cnt}_R \ge m+1$ holds, that is,
whenever $R$ has flipped his bit at least $m+1$ times, $R$ knows the first $m+1$ bits in the sequence,
that is, $X(0)$, $X(1)$ $\dots$ $X(m)$; similarly, whenever  ${cnt}_S \ge
m+1$ holds, that is,
whenever $S$ has flipped his bit $m+1$ times, it must be that $S$ has received acknowledgments that $R$
has received bits $X(0)$, $X(1)$ $\dots$ $X(m)$. (For a formal proof of
this claim, see~\cite{HZ}.
Note that the proof in \cite{HZ} relies essentially on the fact that
messages cannot be reordered.)

As in Stenning's protocol, we need to add some
basic clauses
that ensure that the variables $i_S$ and $i_R$ have the right properties.
These clauses should say that initially $i_S$ is set to $0$,
that $i_S$ is flipped only if $S$ receives the message $i_S$ from $R$, and that, if
infinitely often $R$ receives a message from $R$ equal to $i_S$, then $i_S$ is flipped infinitely often;
similarly, the clauses should say that initially $i_R$ is set to a null value, that $i_R$ is changed only when
$R$ receives a message from $S$ (either equal to $i_R$, if $i_R$ is not null, or not null), and that, if
infinitely often this is the case, then infinitely often $i_R$ is changed. Note that all these clauses
can be easily expressed using the message automata language.
\commentout{
The following clauses, similar in spirit to those used for Stenning's
protocol, do the job.
$$
\begin{array}{l}
\ff{{\cf @} S ~{\cf initially}~ i_S ~=~ 0 }~\oplus \\
\ff{{\cf @}S ~{\cf if}~ kind=local(inc_S)~ {\cf then}~ i_s := 1 -i_S} ~\oplus \\
\ff{{\cf @}S ~kind=local(inc_S)~\cf{ only~ if~} ~ msg(l_{RS}) = i_S}~\oplus \\
\ff{{\cf @}S ~\cf{if ~necessarily~} msg(l_{RS}) = i_S ~ \cf{then ~i.o. ~} kind=local(inc_S)} ~\oplus \\
\ff{{\cf @}S ~{\cf only ~events~ in ~[}rcv(l_{RS}){\cf ] ~affect~} i_S } ~\oplus ~\\
\ff{{\cf @}R ~{\cf initially } ~i_R ~= ~\lambda }~ \oplus\\
\ff{{\cf @}R ~{\cf if}~ kind=local(inc_R)~ {\cf then}~ i_R := f_R(i_R, msg(l_{SR}))} ~\oplus \\
\ff{{\cf @}R ~kind=local(inc_R)~\cf{ only~ if~} ~ (i_R = \lambda) \vee (p_2(msg(l_{SR})) \neq i_R)}~\oplus \\
\ff{{\cf @}R ~\cf{if ~necessarily~}  (i_R = \lambda) \vee (p_2(msg(l_{SR})) \neq i_R)~ \cf{then ~i.o. ~} kind=local(inc_R)} ~\oplus \\
\ff{{\cf @}R ~{\cf only ~events~ in ~[}rcv(l_{SR}){\cf ] ~affect~} i_R },
\end{array}
$$
$p_2(a)$ represents the second element of a pair $a$
and $f_R$ is a function such that $f_R(i,v)=v$ if $i=\lambda$ and $f_R(i,v)=1-i$ otherwise.
}%

\commentout{
We also assume that agents have perfect recall.
We need to make this assumption since we are looking here for
instantiating the knowledge-based
derivation in the previous section, and that derivation relies heavily on the assumption that
the two predicates, one for the sender, and one for the receiver, are stable; the most straightforward way
of ensuring stability we see is to assume perfect recall. Clearly, this means that the
protocol we will derive here is infinite-state, while the original alternating-bit
protocol is finite-state. However, this does not say that a derivation for the original
alternating-bit protocol is not possible in Nuprl. In fact, we could instead assume a third agent
in the system (i.e., the environment), whose state encodes each player's history, and have the sender's and
the receiver's local states be some projection of their history.
}%

We now show that $\tilde{\phi}_S^{kb}$, $\tilde{\phi}_R^{kb}$,
 $\tilde{t}_S$ and $\tilde{t}_R$ chosen as
 above satisfy all the conditions identified during the derivation at the beginning of this section.
We do not give a formal proof here; rather, we present enough details
for the reader to have an understanding of how the proof works.
It is not difficult to see that both
$\ff{\Stable({\tilde{\varphi}}^{kb}_R)}$ and
$\ff{\Stable({\tilde{\varphi}}^{kb}_S)}$ hold, as both ${cnt}_S$ and
${cnt}_R$ can never decrease, and that $\ff
{\Dec({\tilde{\varphi}}^{kb}_R )}$ and $\ff
{\Dec({\tilde{\varphi}}^{kb}_S )}$ also hold. To see that ${\ff
Fair^{kb}_I(\varphi^{kb}_R,\tilde{t}_R,l_{RS})}$ also holds, recall
that $\varphi^{kb}_R$ is defined as $\exists n.~\forall
k<n.~\tilde{\phi}^{kb}_R [ m / k ] \wedge \neg \tilde{\phi}^{kb}_R [
m / n ]$, which is equivalent to $\exists n. \forall
k<n.~({cnt}_R\ge k+1) \wedge ({cnt}_R < n+1)$, that is, $\exists
n.~{cnt}_R = n$, and so $\varphi^{kb}_R$ is always true. By
inspecting the definition of ${\ff
Fair^{kb}_I(\varphi^{kb}_R,\tilde{t}_R,l_{RS})}$, this implies that
${\ff Fair^{kb}_I(\varphi^{kb}_R,\tilde{t}_R,l_{RS})}$ is reduced to
showing that the following holds in all runs $\es$ of the
alternating-bit protocol: $\exists e_R@R\in \es \wedge \forall
e@R\in \es.~\exists e'\in \es. \kind(e')=\rcv(l_{RS}) \wedge
\send(e')\succeq e@R$. In other words, we need to show that, in all
runs of the alternating-bit protocol, some event occurs associated
with $R$, and for all events associated with $R$, such as $R$
receiving a message from $S$, there will be a subsequent message
sent by $R$ to $S$ and received by $S$. This is clearly true for the
alternating-bit protocol. Similarly, ${\ff
Fair^{kb}_I(\varphi^{kb}_S,\tilde{t}_S,l_{SR})}$ is reduced to the
condition $\exists e@S \in \es \wedge \forall e@S\in \es. ~\exists
e'.~ \kind(e')=\rcv(l_{SR}) \wedge \send(e')\succeq e@S\in \es$,
which basically says that some event associated with $S$ occurs and
that, whenever $S$ receives a message from $R$, there will be a
subsequent message sent by $S$ to $R$ and received by $R$. Again,
this is true for all runs of the alternating-bit protocol;
that is, $\varphi^{kb}_S$, like $\varphi^{kb}_S$, is equivalent to the
formula $\true$ in all runs of the system corresponding to the
alternating-bit protocol.

The formula
$\ff{\Imp(\tilde{\phi}_S^{kb},\tilde{\phi}_R^{kb})}$ is equivalent in
this case to the following formula:
$$\ff{\forall n}.~\forall e_S@S\in \es.~ (({cnt}_S \ge n+1) ~\before~e_S) \rimp
\exists e_R \succ e_S @R \in \es.~(({cnt}_R \ge n+1)~\after~e_R).$$
This says that if the sender has flipped his bit at
least $n+1$ times before he sends a message to $R$, upon receiving
that message $R$ will have flipped his bit at least $n+1$ times, as
well. In fact, we can see that with the alternating-bit protocol
(that is, with the enforced semantics for $i_S$ and $i_R$),
if
$S$ has flipped his bit exactly $k$ times before he sends a message
to $R$, and if that message is received by $R$, then, when $R$
receives this message, either $R$ has already flipped his bit
exactly $k+1$ times and discards this message, or $R$ has flipped
his bit $k$ times, $R$ does not discard this message and flips his
bit one more time, ensuring $R$ will have been flipped his bit $k+1$
times after receiving this message.

The formula $\ff{\Rcv(\tilde{\phi}_S^{kb}, \tilde{\phi}_R^{kb}, l_{RS})}$
is equivalent to
$$
\begin{array}{l}
\ff{\forall e_S@S.~(\kind(e_S)=\rcv(l_{RS})) \rimp } \\
\quad \forall n. \ff{((\forall k\le n.~ (({cnt}_R \ge
k+1)~\before~\send(e_S))) \rimp
(({cnt}_S \ge n+1)~\after~e_S))}.
\end{array}
$$
This formula basically says that if $R$ has flipped his bit at least $n+1$
times before sending a message to $S$, and $S$ receives this
message, then $S$ will have flipped his bit at least $n+1$ after
seeing this message. We leave it to the reader to check that this is
true for the runs of the alternating-bit protocol
(again, based on the enforced semantics for $i_S$ and $i_R$).

Finally, the formula $\ff{\Rcv(\tilde{\phi}_R^{kb}, \tilde{\phi}_S^{kb},
l_{SR})}$ is equivalent to
$$
\begin{array}{ll}
\ff{\forall e_R@R.} & \ff{(\kind(e_R) = \rcv(l_{SR})) \rimp }\\
& \ff{\forall n.
~(((\forall k<n.~ {cnt}_S\ge k+1 \wedge {cnt}_S < n+1) ~\before~\send(e_R)) } \\
& \quad \quad \rimp \ff{(({cnt}_R \ge n+1)~after~e_R))}.
\end{array}
$$
This formula says that if $S$ has flipped his bit exactly $n$  times
before he sends
a message to $R$, and $R$ receives this message, then after
receiving this message $R$ will have flipped his bit at least $n+1$
times. This easily follows from the argument made
above that $\Imp(\tilde{\phi}_S^{kb}, \tilde{\phi}_R^{kb})$ holds.
It follows that all the conditions that we identified for deriving a standard
program from a knowledge-based program are satisfied.  Thus,
if messages are not reordered, the specification for the
sequence-transmission problem is satisfied by the standard program
$$Fair\mbox{-}Pg^{kb}( \phi^{kb}_S, \tilde{t}_S, l_{SR}, a_S) \oplus
Fair\mbox{-}Pg^{kb}_I( \phi^{kb}_R, \tilde{t}_R, l_{RS}, a_R).$$
Since, as we showed above, both $\phi_S^{kb}$ and $\phi_R^{kb}$ are
always true for our particular choices of
$\tilde{\phi}_S^{kb}$, $\tilde{\phi}_R^{kb}$, $\tilde{t}_S$, and
$\tilde{t}_R$, this becomes
$$
\begin{array}{l}
\ff{{\cf @}S ~kind=local(a_S) ~{\cf only ~if ~} true } ~\oplus \\
\ff{{\cf@}S~ {\cf if}~ kind=local(a_S) ~{\cf then}~ msg(l_{SR})~ := ~ \< x_S, i_S \> } ~\oplus  ~\\
\ff{{\cf @}S ~{\cf only ~events~ in ~[}a_S {\cf ] ~affect~} msg(l_{SR}) } ~\oplus ~\\
\ff{{\cf @}S~{\cf if~ neccessarily~ }true ~{\cf then~ i.o.~} kind=local(a_S)} ~\oplus ~\\
\ff{{\cf@}R ~kind=local(a_R) ~{\cf only ~if~} true }~ \oplus ~\\
\ff{{\cf@}R~ {\cf if}~ kind=local(a_R)~ {\cf then}~ msg(l_{RS}) ~:=~ i_R } ~\oplus ~\\
\ff{{\cf@}R ~{\cf only ~events ~in ~[ }a_R {\cf] ~affect}~ msg(l_{RS}) }~\oplus ~\\
\ff{{\cf@}R~{\cf if ~neccessarily~} true ~{\cf then~ i.o.}~ kind=local(a_R)}.
\end{array}
$$
Note that this program indeed corresponds to the alternating-bit protocol.

\section{Conclusion and Future Work}\label{sec:conc}
We have shown that the mechanism for synthesizing programs from
specifications in Nuprl can be extended to knowledge-based programs
and specifications,
Moreover, we have shown that axioms much in the spirit of those used
for
standard programs can be used to synthesize knowledge-based programs as well.
We  applied this methodology to the analysis of the
sequence-transmission problem, and showed that the knowledge-based
programs proposed by
Halpern and Zuck for solving the STP problem can be synthesized in
Nuprl.
We also sketched an approach for deriving standard programs that
implement the knowledge-based programs that solve the STP.
A feature of our approach is that the extracted standard programs
are closer to the pseudocode that designers write,
and can be translated into running code.

There has been work on synthesizing both standard programs and
knowledge-based programs from knowledge-based specifications.
In the case of synchronous systems with only one process, Van der
Meyden and Vardi \cite{meyden-synthesis} provide
a necessary and sufficient condition for a certain type of knowledge-based
specification to be realizable, and show that, when it holds, a
program can be extracted that satisfies the specification. Still
assuming a synchronous setting, but this time allowing
multiple agents,  Engelhardt, van der Meyden, and Moses
\cite{EMMFoss,EMM01} propose a refinement calculus in which one
can start with an epistemic and temporal specification and use
refinement rules that eventually lead to standard formulas. The
refinement rules annotate formulas with preconditions and
postconditions, which allow programs to be synthesized from the leaf
formulas in a straightforward way. A search up the tree generated in
the refinement process suffices to build a program that satisfies
the specification.
The extracted programs are objects of a programming language that
allows concurrent and sequential executions, variable assignments,
loops and conditional statements.

We view our method for synthesizing programs from knowledge-based
specifications
as
an alternative to this approach. As in the Engelhart et
al.~approach, the Nuprl programs that we extract are close to programs in
standard programming languages. Arguably, distributed I/O message
automata are general enough to express most of the distributed
programs of interest when communication is done by message passing.
Our approach has the additional advantage of working in asynchronous
settings. \commentout{ Clearly, alternative choices for the type of
programs to be synthesized exists. For example, programs can be
expressed in a standard process algebra, like CSP or CCS. Such  a
choice has a number of benefits: both CSP and CCS are abstract
calculi of sequential and concurrent processes in which complex
specifications can be expressed in an elegant way, and there is a
well-developed theory of satisfiability.
On the other hand, it is important to synthesize programs that can
be relatively naturally translated to running code; in this respect,
recent results on translating IO message automata to Java code
sustain the idea that our approach is viable. We leave the
comparison of the expressive power of these process languages and
their suitability to program synthesis to future research. }

A number of questions, both theoretical and more applicative, still
remain open. While synthesis of distributed programs from epistemic
and temporal specifications is not computable in general, recent
results \cite{meyden_syn05} show that, under certain assumptions
about the setting in which agents communicate, the problem is
computable. It would be worth understanding the extent to which these
assumptions apply to our setting.
Arguably, to prove a result of this type, we need a better
understanding of how properties of a number of knowledge-based programs
relate to
the properties of their composition; this would
also
allow us to prove stronger composition rules than the one presented
in Section~\ref{sec:kb}.
As we said, we believe that the approach that we sketched for
extracting a standard
program from the knowledge-based specification for the STP problem can be
extended into a general methodology. As pointed out by Engelhart et
al., the key difficulty in extracting standard programs from
abstract specifications is in coming up with good standard
tests to replace the abstract tests in a program. However, it is
likely that, by reducing the complexity of the problem and focusing
only on certain
classes of knowledge-based specifications, ``good'' standard tests can be more
easily identified.
\section*{Acknowledgements}
We would like to thank Richard Eaton from the Nuprl group for making the
Nuprl lemma corresponding to our proof of the sequence-transmission
problem available online.

\bibliographystyle{plain}
%
%
%
%
%
%
%
%
%
%
%
%
%
%
%
\bibliography{synthesis_submission}

\end{document}